\documentclass[fleqn,usenatbib]{mnras}

\usepackage{newtxtext,newtxmath}

\usepackage[T1]{fontenc}
\usepackage{ae,aecompl}

\usepackage{graphicx}	
\usepackage{amsmath}	
\usepackage{amssymb}	
\usepackage{mathtools}
\usepackage{cleveref}
\usepackage[dvipsnames]{xcolor}

\hypersetup{draft}

\title[Age Demographics of the Milky Way]{Age Demographics of the Milky Way Disk and Bulge}

\author[J. Grady et al.]{
J. Grady,$^{1}$\thanks{E-mail: jrg71,vasily,nwe@cam.ac.uk}
V. Belokurov,$^{1}$
N.W. Evans$^{1}$
\\
$^{1}$Institute of Astronomy, University of Cambridge, Madingley Road, Cambridge, CB3 0HA, United Kingdom\\
}

\date{Accepted XXX. Received YYY; in original form ZZZ}

\pubyear{2019}

\begin{document}
\label{firstpage}
\pagerange{\pageref{firstpage}--\pageref{lastpage}}
\maketitle

\begin{abstract}
We exploit the extensive $Gaia$ Data Release 2 set of Long Period Variables to select a sample of Oxygen-rich Miras throughout the Milky Way disk and bulge for study. Exploiting the relation between Mira pulsation period and stellar age/chemistry, we slice the stellar density of the Galactic disk and bulge as a function of period. We find the morphology of both components evolves as a function of stellar age/chemistry with the stellar disk being stubby at old ages, becoming progressively thinner and more radially extended at younger stellar ages, consistent with the picture of inside-out and upside-down formation of the Milky Way's disk. We see evidence of a perturbed disk, with large-scale stellar over-densities visible both in and away from the stellar plane. We find the bulge is well modelled by a triaxial boxy distribution with an axis ratio of $\sim [1:0.4:0.3]$. The oldest of the Miras ($\sim$ 9-10 Gyr) show little bar-like morphology, whilst the younger stars appear inclined at a viewing angle of $\sim 21^{\circ}$ to the Sun-Galactic Centre line. This suggests that bar formation and buckling took place 8-9 Gyr ago, with the older Miras being hot enough to avoid being trapped by the growing bar. We find the youngest Miras to exhibit a strong peanut morphology, bearing the characteristic X-shape of an inclined bar structure.
\end{abstract}




\section{Introduction}
The Milky Way disk teems with complex structure and kinematic features
that portray the turbulent history of formation, interaction and
evolution. The detailed mechanisms of disk formation remain to be
worked out, but the general picture is clear. Early infall of halo gas
provides material that cools and settles into a disk in the inner
regions of the galaxy. Gas accreted at later times falls onto the
outer regions providing an 'inside-out' picture for the formation of
the stellar disk \citep[see e.g.][]{Larson_1976,Kepner_1999,Mo10}.

In a now classic paper, \citet{Gilmore_1983}  studied
star counts toward the Galactic South pole and
found the vertical stellar profile to be well modelled by a double exponential; a vertically extended component with a scale height of $\sim 1$ kpc and a thinner disk with scale height $\sim$ $0.3$ kpc, leading to the notion of distinct thin and thick discs both constituting the Galaxy's stellar disk More recent studies of the disk density profile have also found a distinction between the thin and thick discs in terms of scale lengths and scale heights \citep[see e.g.][]{Ojha_2001,Yoachim_2006,Juric_2008} with the distinction being purely geometric; the thick disk is both vertically and radially more extended than the thin disk. Further, such thick disk components are a common feature in other disk galaxies \citep[see e.g.][]{Yoachim_2006,Pohlen_2007,Comeron_2012}. The Galaxy's thick disk component has generally been observed as older \citep{Liu_2000,Kilic_2017} and kinematically hotter \citep{Soubiran_2003,Yoachim_2006} than the thin disk. 

Modern spectroscopic studies \citep[see e.g.][]{Mishenina_2004,Reddy_2006,Adibekyan_2011,Adibekyan_2012,Hayden_2015} have revealed a chemical dichotomy within the disk. There exists a population of stars enhanced in $\left [ \alpha/\textup{Fe} \right ]$ and deficient in $\left [ \textup{Fe} / \textup{H} \right ]$ comprising an older component. Conversely, there is a younger population enhanced in $\left [ \textup{Fe} / \textup{H} \right ]$ and deficient in $\left [ \alpha/\textup{Fe} \right ]$. The work of \citet{Bovy_2012} investigated the spatial structure of the disk through a sample of G-dwarfs from the \textcolor{black}{Sloan Extension for Galactic Understanding and Exploration} (SEGUE) survey. Having both $\left [ \alpha/\textup{Fe} \right ]$ and $\left [ \textup{Fe}/\textup{H} \right ]$ information, they divided their sample into mono-abundance sub populations. The density profile of each sub-population was well modelled by a single exponential in both the radial and vertical directions. In this analysis, the scale length (height) decreases (increases) for older sub-populations, in stark contrast to the geometric thick and thin disk. Thus, it has now become common to refer to the $\alpha$-rich and $\alpha$-poor discs, in favour of the traditional thick and thin discs. 

Separately, and core to the makeup of the central galaxy, is the Galactic bulge. The bulge has garnered much interest in order to ascertain the origin of this structure. Two main theories exist with one positing the bulge formed as a consequence of successive mergers at early stages in the Milky Way's history. The competing scenario sees the bulge form secularly, born out of natural dynamical evolution of the stellar disk. These differing formation scenarios yield distinct structures that, following the convention of \citet{Kormendy_2004}, are called classical bulges and pseudo-bulges respectively. Consequently, the structural nature of the Milky Way bulge has been much studied to ascertain its exact nature. Utilising the near infrared imaging of the COBE satellite, \citet{Dwek_1995} confirmed the global structure of the bulge to be 'boxy', complementing the barred structure observed by \citet{Weiland_1994} in the same data set. Subsequent analysis of Red Clump (RC) stars towards the Galactic centre by \citet{McWilliam_2010} revealed a split in the stellar luminosity function, interpreted as the signature of an X-shaped bulge/boxy-peanut (BP) bulge. Such a morphology is a clear indicator of a pseudo-bulge component central to the galaxy, formed out of the buckling instability known to affect barred structures and supported by resonant banana orbits~\citep{Wi16}. \textcolor{black}{Recent mappings of the inner and outer bulge regions with RC stars from the Vista Variables in the Via Lactea (VVV) survey} by \citet{Wegg_2013} and \citet{Wegg_2015} have further confirmed the B/P nature of the inner bulge that proceeds to extend and flatten into a barred structure into the disk, both structures oriented at an angle of $\sim 27^{\circ}$. Whilst this interpretation has been contested with \citet{Lopez_2019} claiming that the split RC is a result of two distinct stellar populations, the spectacular WISE imaging of \citet{Ness_2016} reveals the X-shape clearly. The proper motion data from VVV as analyzed by \citet{Sa19} show differential motion between the peaks, which argues strongly against population effects causing the X-shape. This picture is complicated, however, by claims of ancient stellar populations of RR Lyrae and Mira forming a distinct central component, with no indication of tracing the position angle of the bar \citep[see e.g.][]{Dekany_2013,Catchpole_2016,Gran_2016,Prudil_2019}, suggestive of a composite nature to the bulge. 


This work conducts a similar analysis of the MW disk and Bulge on the hitherto unexploited Mira sample provided by the $Gaia$ Data Release 2 (DR2) \citet{Gaia_2016,Gaia_2018} and detailed in \citet{Gaia_LPVs}. Miras are long period variables (LPVs) residing on the asymptotic giant branch (AGB). They pulsate with periods ranging from 100 - 1,000 days and typically have pulsation amplitudes $\geq$ 2.5 magnitudes in the visible band. They obey a well defined period-luminosity (PL) relation \citep{Glass_1982} in the near-infrared making them useful distance indicators. Further, they are intrinsically very bright and easily visible in the Magellanic Clouds \citep{Deason_2017}, \textcolor{black}{ M31 and M33 \citep{An_2004} and have even been observed in NGC1559 \citep{Huang_2019}.} \textcolor{black}{A further important characteristic of Miras is their ability to act as chronometers. The work of \citet{Feast_2006} demonstrated that the pulsation period of Miras is inversely correlated with their vertical velocity dispersion, implying longer period Miras are in fact younger than their short period counterparts. A sample of age estimates is provided in \citet{Feast_2009}. Further, \citet{Wyatt_1983} also computed Mira age estimates, as a function of their pulsation period, by estimating the main sequence lifetime of a Mira sample based on their luminosity; the time spent on the AGB track being negligible in comparison to their main sequence life times. Both methods yield age estimates for Miras that are consistent within $\sim 1$ Gyr}. The study of  Miras in Galactic globular clusters by \citet{Feast_2000} highlighted the existence of a period-metallicity relation for the O-rich Miras; stars with longer pulsation periods bear higher metallicity values. Thus, for Mira variables, the age, the abundance $\left [ \alpha/\textup{Fe} \right ]$ and the metallicity $\left [ \textup{Fe}/\textup{H} \right ]$ are encoded in the pulsation period. 

\section{Data}

We make use of the substantial DR2 LPV data set provided by the $Gaia$ consortium~\citet{Gaia_LPVs}. Specifically, we select all stars from the \texttt{vari\_long\_period\_variable} table for which time series information is available -- most crucially, the pulsation period and hence distance estimates. We then cross-match with the Two Micron All Sky Survey (2MASS) point source catalogue to obtain near infrared (NIR) $JHK_{s}$ photometry yielding 89,555 sources. The cross-match was carried out using a 1 arcsec aperture. We correct for extinction using the map of \citet{SFD_map} and \textcolor{black}{extinction coefficients} of $\mathcal{R}_{J}$ = 0.72, $\mathcal{R}_{H}$ = 0.46 and $\mathcal{R}_{K_{s}}$ = 0.306 from \citet{Yuan_2013}.
\begin{figure}
\centering
	\includegraphics[width=\columnwidth]{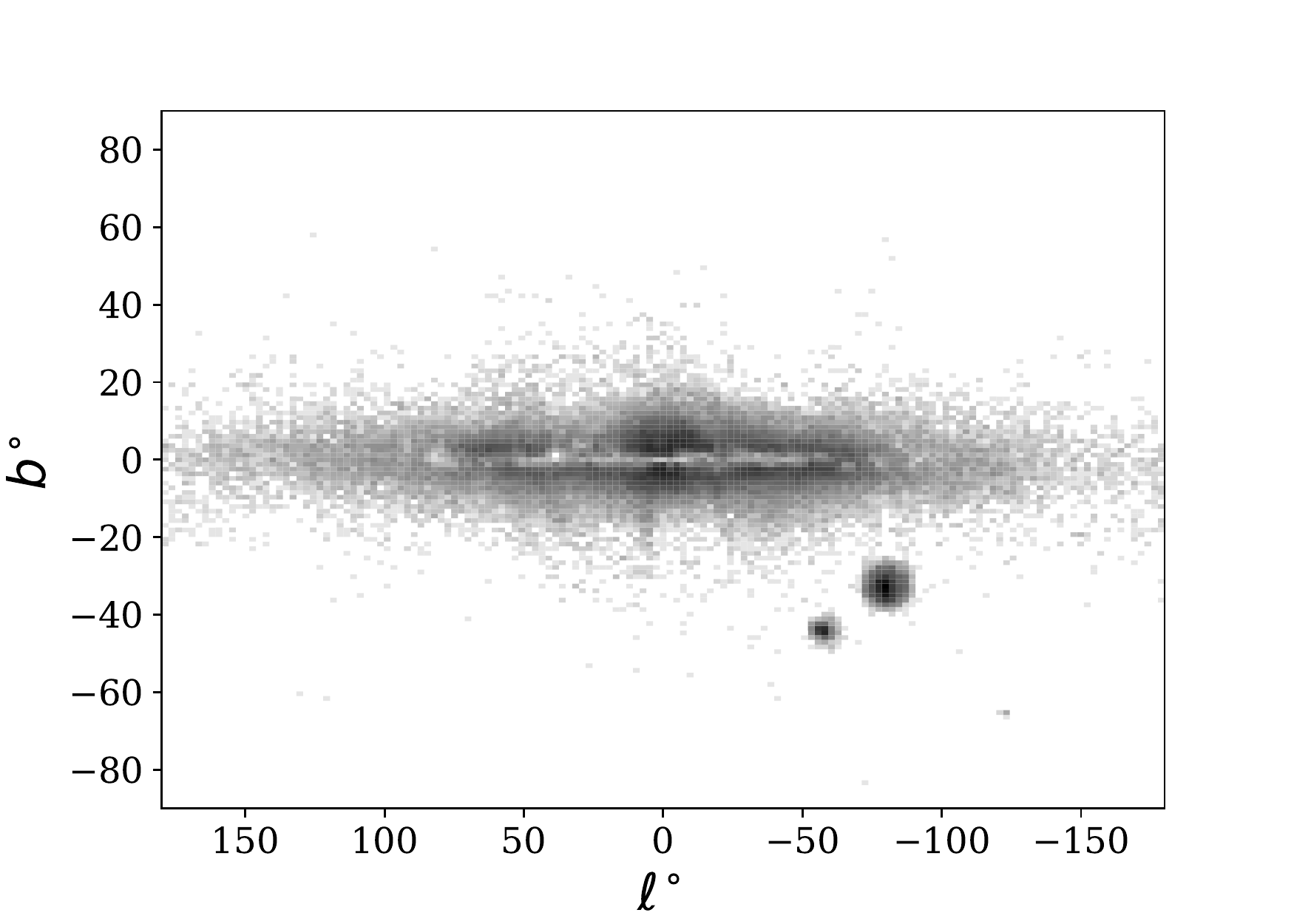}
    \caption[width=\columnwidth]{Logarithmic number density of the $Gaia$ DR2 + 2MASS LPVs in Galactic coordinates. We clearly see the Magellanic Clouds as well as the Sgr dwarf galaxy. Otherwise, the majority of the LPV population is confined to the plane of the Galaxy.}
\label{fig:lb}
\end{figure}
\subsection{Selecting Miras}
\label{sec:selections} 

\begin{figure}
\centering
	\includegraphics[width=\columnwidth]{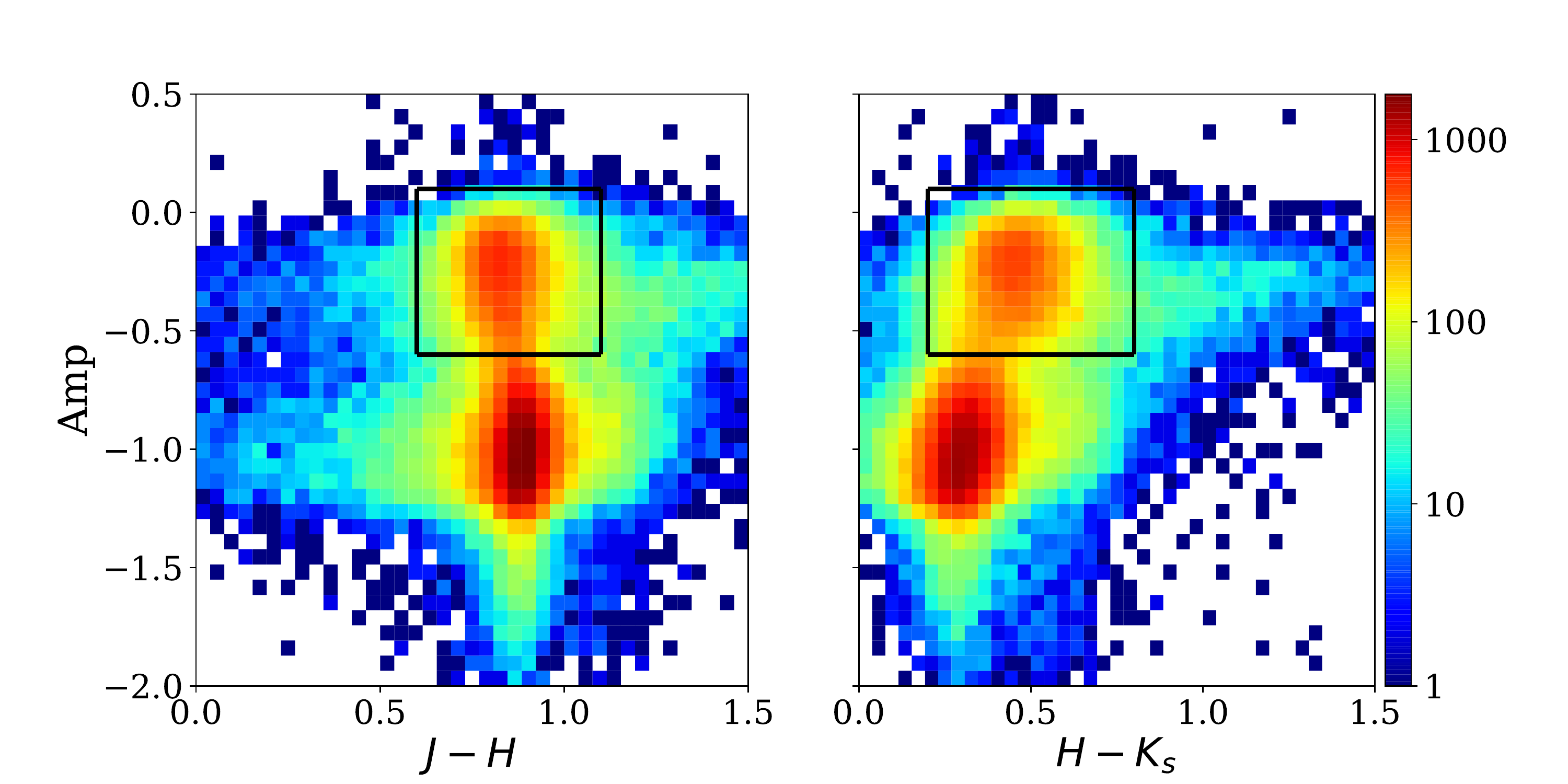}
    \caption[width=\columnwidth]{The logarithmic number density of stars in amplitude colour space for de-reddened $J-H$ and $H-K_{s}$. The Amp term defined in Eq.~\ref{eq:amp} measures the amplitude of variability of the LPVs. We immediately observe a segregation between the high amplitude Miras and the lower amplitude SRVs in this projection. Solid black lines denote our selection boundaries.}
\label{fig:cuts}
\end{figure}

Following the methods of \citet{Vasily_2017}, we define a variability amplitude parameter as:
\begin{equation}
\textup{Amp}  = \textup{log}_{10}   \left (\sqrt{N_{\textup{obs}}}  \: \frac{\sigma_{\overline{I_{G}}}}{\overline{I_{G}}}  \right ) 
\label{eq:amp}
\end{equation}
where $\sigma_{\overline{I_{G}}}$ and $\overline{I_{G}}$ are the mean
flux error and mean flux in the $Gaia$ $G$ band respectively, whilst
$N_{\textup{obs}} $ is the number of observations taken in the
$G$ band. Miras are high amplitude pulsators that have long been known
to lie on precise period-luminosity (PL) sequences in the $JHK_{s}$
system \citep[see e.g.][]{Glass_1982,Feast_1989}. In
Fig.~\ref{fig:cuts} we plot the number density of our $Gaia$ DR2 +
2MASS LPVs in amplitude - colour space where we observe some important
features. The bulk of stars inhabit two regions in amplitude space:
there is a high amplitude population, and one of lower
amplitude. Further, these two populations are largely confined to narrow regions of colour. Beyond this, we see a smearing toward redder and bluer colours. The blue-ward extensions are sources with
large colour excess values lying in heavily extincted regions. The
sources flaring out to large red colours are the Carbon-rich (C-rich)
Miras. Oxygen-rich (O-rich) and C-rich Miras are known to separate in
their near infrared (NIR) colours due to their differing circumstellar compositions
\citep[see e.g.][]{Feast_1982,Glass_1982,Soszynski_2009,Whitelock_2006}. The dusty outskirts of the C-rich Miras may impede our ability to obtain accurate distance estimates to them. \citet{Whitelock_2006b} note that whilst C-rich Miras are observed to lie on the same PL sequence as their counterpart, the O-rich Miras, they do so with higher uncertainty. Further, they can be difficult to separate from C-rich Semi Regular variables (SRV) and overtone pulsators and thus selection of this subset is prone to contamination. The O-rich Miras have less spread in their NIR colours and are identified as the stars bound by the black lines in
Fig.~\ref{fig:cuts}. We select these stars as our sample, which we will henceforth simply refer to as Miras, leaving us with 24,533 stars.
We employ the Mira period-luminosity (PL) relation of \citet{Yuan_2017}, with a minor calibration correction as reported in \citet{Grady_2019}, to obtain:
\begin{equation}
M_{K_{s}} = -6.90 - 3.77\left ( \textup{log}P -2.3 \right ) -2.23\left ( \textup{log}P -2.3 \right )^{2} -0.17
\label{eq:pl}
\end{equation}
with $M_{K_{s}}$ being the absolute magnitude of a Mira and $P$ its pulsation period in days. Once we have computed the luminosity of each Mira using Eq.~\ref{eq:pl}, we can directly estimate the heliocentric distance via the standard relation:
\begin{equation}
    \textup{log}\left ( \frac{D}{\textup{kpc}} \right )  = \frac{ K - M_{K_{s}}}{5} - 2
    \label{eq:D}
\end{equation}
\subsection{Miras in the Large Magellanic Cloud}

\begin{figure}
\centering
	\includegraphics[width=\columnwidth]{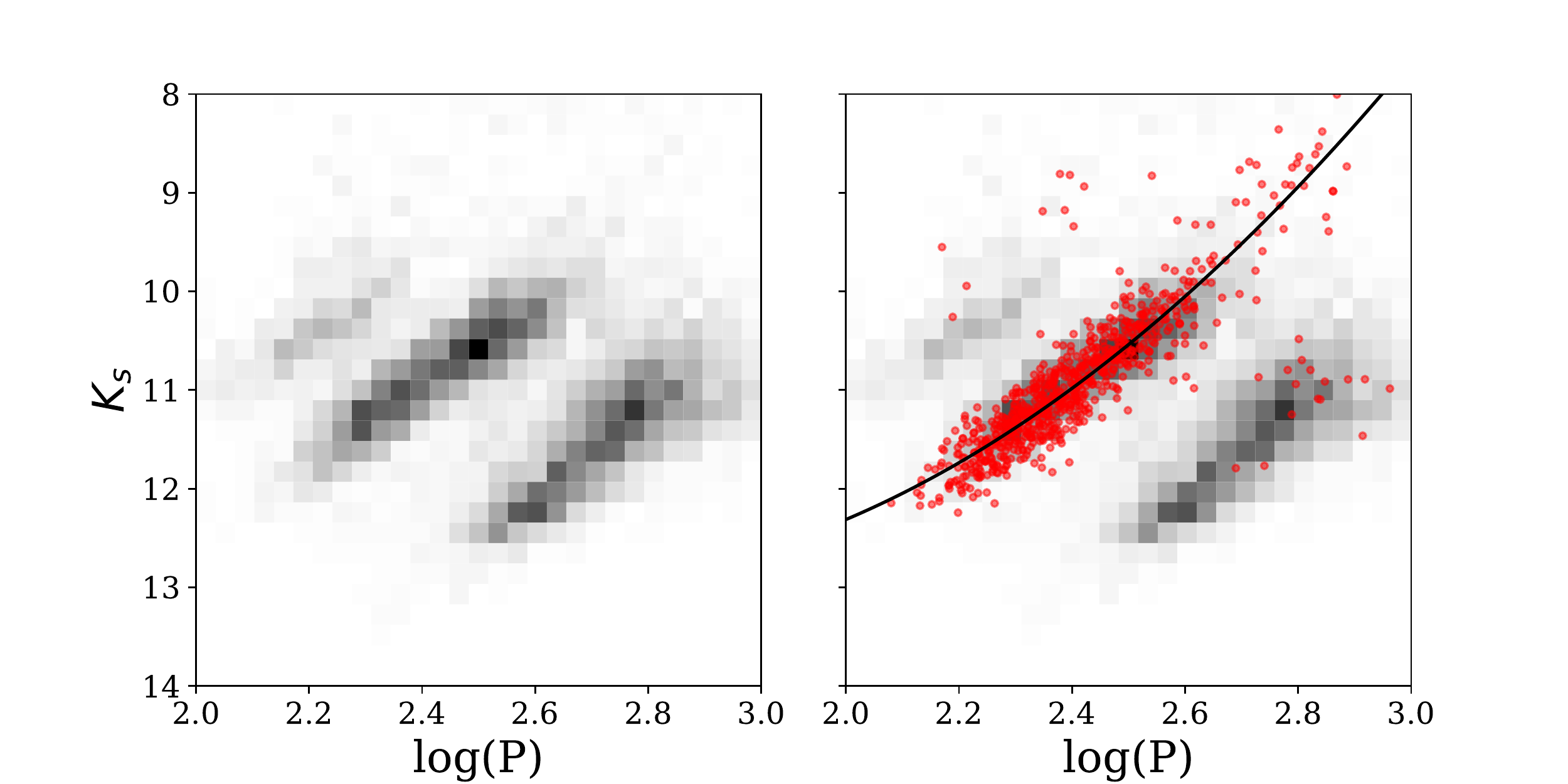}
    \caption[width=\columnwidth]{\textit{Left}: Number density of the $Gaia$ DR2 + 2MASS LPVs selected within a $15^{\circ}$ aperture of the LMC are shown in period magnitude space. Values for $K_{s}$ have been corrected for extinction. Three distinct PL sequences are seen with the middle of these identified as the Mira sequence. The upper and lower sequences are predominately populated by SRVs. \textit{Right}: We overlay our Mira candidates as red markers. These stars are selected from the amplitude - colour cuts shown in Fig.~\ref{fig:cuts}. We see that our selection acts well to pick out the Miras for which we can obtain accurate distance estimates. The solid black line shows our PL relation of Eq.~\ref{eq:pl} adjusted to the LMC's distance modulus.}
\label{fig:lmc}
\end{figure}

\textcolor{black}{We test the efficacy of our selection method by analysing the $Gaia$ DR2 + 2MASS LPVs towards the Large Magellanic Cloud (LMC). Selecting stars within a $15^{\circ}$ aperture of the LMC, we show their distribution in $K_{s}$ magnitude - period space in the left panel of Fig.~\ref{fig:lmc}. Three main sequences are prominent with the middle of these being the high amplitude Mira sequence that we wish to select from \citep[see][]{Wo99}. On imposing the selection of Fig.~\ref{fig:cuts} upon the LMC sample, we select stars mainly from the Mira sequence, as demonstrated in the right hand panel of Fig.~\ref{fig:lmc}. We overlay our PL relation of Eq.~\ref{eq:pl}, corrected for the LMC distance modulus of \citet{Elgueta_2016} as a solid black line. There is a degree of contamination, as shown by the stars deviating from the Mira sequence for which we are unable to obtain accurate distances. We estimate the contaminant fraction by taking the ratio of the number of Miras deviating from our PL relation by > 0.5 magnitudes with the total number of selected LMC Miras. This yields a contamination estimate of $\sim$ 7$\%$. Utilising the sample of LMC LPVs of \citet{Soszynski_2009}, we estimate the purity of our O-rich Mira sample to be $72 \%$ with the major contaminant being high amplitude stars classified as SRVs by OGLE. We recover $47 \%$ of the OGLE-III O-Miras in the cross-match, a completeness value consistent with that of the full $Gaia$ DR2 LPV catalogue with respect to the OGLE-III LPVS towards the Magellanic Clouds \citep[see][]{Gaia_LPVs}.}
\subsection{Galactic Sample}

Upon selecting our full sample of Mira, we assign distances by the procedure outlined and compute the Galactocentric $(X,Y,Z)$ coordinates for each star. We then remove stars lying towards the Magellanic Clouds by excising regions on the sky located within $15^{\circ}$ and $10^{\circ}$ of the LMC and SMC respectively. We further restrict our analysis to Miras in the period range of $100-400$ days, beyond which there may be increased scatter about the PL relation, producing a sample size of 21,149 O-rich Miras. \textcolor{black}{We choose 400 days as our upper limit as the $Gaia$ DR2 data sample is based on data collected from a 22 month period. Thus, the number of candidate Mira with periods exceeding 400 days in $Gaia$ DR2 is highly depleted \citep[see][]{Gaia_LPVs}}.

Our selection criteria utilises an amplitude criterion defined by the $Gaia$ photometry despite the fact that light curve amplitudes are supplied for the LPVs in $Gaia$ DR2. We made a similar selection to those of Fig.~\ref{fig:cuts} using the \texttt{range\_mag\_g\_fov} amplitude column in the \texttt{vari\_time\_series\_statistics} table from $Gaia$ DR2. This yielded a sample of Miras with an equivalent contamination estimate, but a sample size $\sim 15\%$ smaller than the one we have obtained here. Thus, our rationale for using the amplitude definition of Eq.~\ref{eq:amp} is simply down to the advantage of a larger data set for equivalent contamination. 

\begin{figure*}
\centering
	\includegraphics[width=\textwidth]{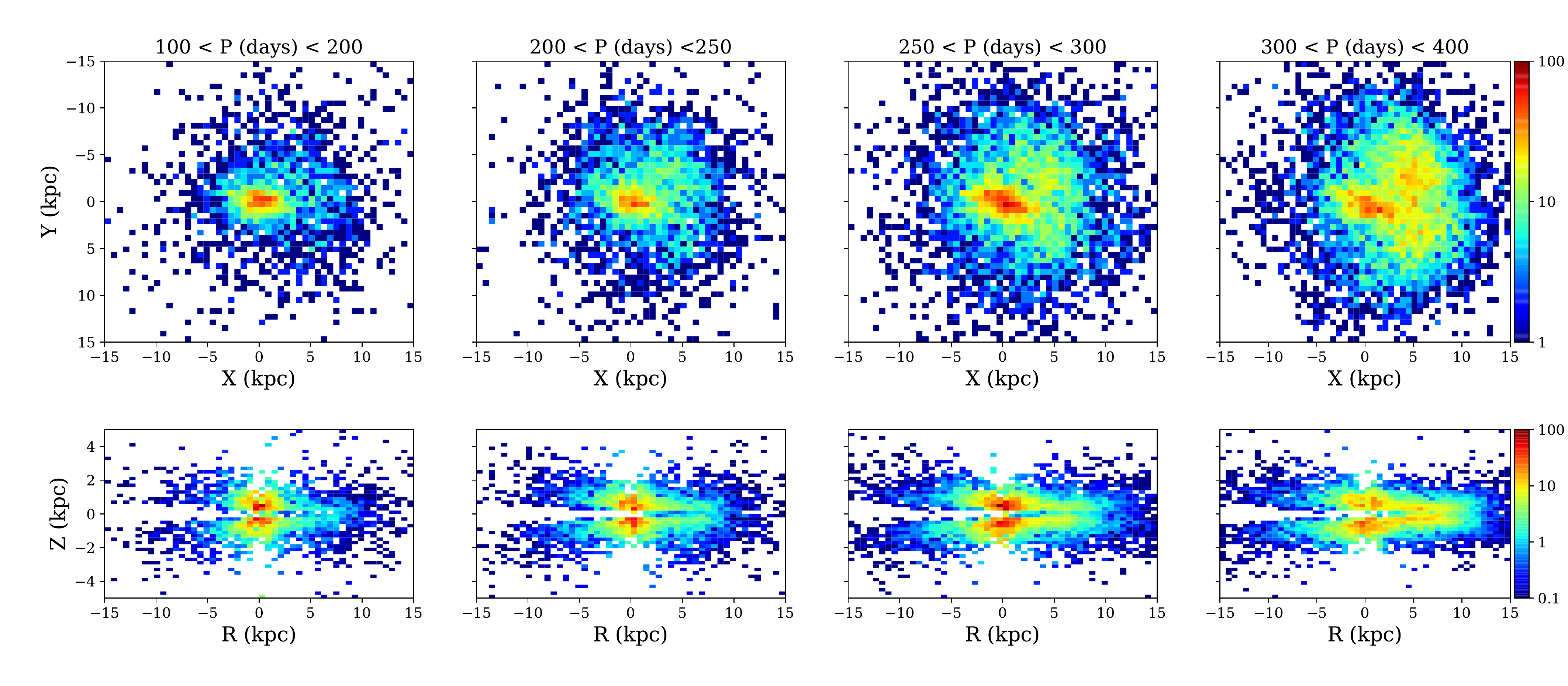}
    \caption[width=\textwidth]{We show the spatial distribution of our Mira sample subdivided into four periods bins. \textit{Top}: Galactocentric $X-Y$ projection of our sample with the Solar position at $(8.3,0,0)$. \textit{Bottom}: Galactocentric cylindrical $(R,Z)$ projection where we see most of the stars vertically confined to within $2$ kpc of the Galactic plane. Each column represents Miras in a different period range evolving from an older/metal poorer population on the left to a younger/metal richer population on the right. The bulge/bar is visible throughout, with the disk population becoming more populous with increasing Mira period.}
\label{fig:XY_RZ}
\end{figure*}

The distribution of stars in Fig.~\ref{fig:XY_RZ} shows clear evolution as a function of period. The bulge grows in scale from low to high period with a distinct tilt being apparent at higher periods. The bulge dominates the Mira population at low periods, as expected for an older, metal poorer stellar population. For the longer period Miras, we see an extended bar stretching out to $\sim 5$ kpc either side of the galaxy. With regards to the disk, there is a plentiful stellar population local to the Sun at high periods, these being the young, low latitude, nearby Miras. Their vertical extent is restricted and they extend to beyond $\sim 10$ kpc radially. The lower period stars are less restricted to the plane, demonstrating a fluffier disk component for the older/metal poorer Miras. At the lowest periods, we observe the disk to be radially truncated in comparison to the higher period stars. This is suggestive of a progression from a thin, radially extensive disk at young ages to a thicker, stubbier disk at older stellar ages. 
\section{Stellar Density Profiles}
\subsection{Modelling the Galactic Disk and Bulge}
\label{sec:model}

Each star is modelled as a realisation of a Poisson process at its spatial position. Thus, for a given density model, the likelihood has the general form:
\begin{equation}
    \textup{ln} \: \mathcal{L} = -\int_{V} \rho\left ( \textbf{r}  | \mathbf{\Theta} \right ) S\left ( \textbf{r} \right ) dV + \sum_{i=1}^{N} \textup{ln} \: \rho\left ( \textbf{r}_{i}  | \mathbf{\Theta} \right )
    \label{eq:lnL}
\end{equation}
where $\rho\left ( \textbf{r} | \mathbf{\Theta} \right )$ is the
number density defined by model parameters $\mathbf{\Theta}$ and $\textbf{r}$
the Galactocentric position vector. The
integral part of the likelihood is a normalising term which equates to
the number of Miras within the total volume $V$ and the right hand
summation is conducted over all stars bound by the volume. Any general
selection function $S\left ( \textbf{r} \right )$ depending on the
spatial coordinates can in principle be incorporated. In our application, the
selection function is highly non-trivial being a convolution of the
individual selection functions of 2MASS and $Gaia$, the latter being unknown. We remark that the distribution of LPVs in
Fig.~\ref{fig:lb} is encouragingly smooth. That is, there is no
obvious spatial dependence on detection efficiency through the Galaxy
for the $Gaia$ LPVs. We therefore only account for the footprint over
which we can observe our Miras, as detailed later in this section.

The presence of a strong bulge/bar in the Miras, as seen in Fig.~\ref{fig:XY_RZ}, motivates us to adopt a two component model to
describe the data: a disk and a bar/bulge. We model the disk component by a simple double exponential function in Galactocentric cylindrical polar coordinates $(R,Z)$. The bulge/bar component is modelled as a triaxial boxy Gaussian distribution as in \citet{Dwek_1995}. Explicitly, these are:
\begin{subequations}
\begin{align}
        \label{eq:disc_density}
        \rho_{\textup{disk}}(R,Z) &= \rho_{0}^{\textup{disk}}\,   \textup{exp}\,\left (  -R/R_{s} \right ) \textup{exp}\,\left (  -|Z|/Z_{s} \right ) \\
        \label{eq:bar_density}
        \rho_{\textup{bulge}}(x,y,z) &= \rho_{0}^{\textup{bulge}}\,   \textup{exp}\,\left (  {-0.5\,m^{2}(x,y,z)} \right )
\end{align}
\end{subequations}
where $R_{s}$ and $Z_{s}$ denote the scale length and scale height of the Galactic disk. The terms $\rho_{0}^{\textup{disk}}$ and $\rho_{0}^{\textup{bar}}$ are the central number densities determined by the total number of stars included in the normalising volume integral of Eq.~\ref{eq:lnL}. The term $m$ in Eq.~\ref{eq:bar_density} is computed as:
\begin{equation}
    m = \left [\left( \left ( \frac{x}{x_{0}} \right )^{2} + \left ( \frac{y}{y_{0}} \right )^{2} \right)^2 + \left ( \frac{z}{z_{0}} \right )^{4} \right ]^{\frac{1}{4}}
    \label{eq:r}
\end{equation}
where $(x_{0},y_{0},z_{0})$ represent the principal semi-axes of the
bar. The Cartesian coordinates $(x,y,z)$ are aligned along the major,
intermediate and minor axes of the bar. Fig.~\ref{fig:XY_RZ} shows
the major axis of the bar to be misaligned with the line of sight from
the Sun to the Galactic Centre (GC). We transform from Galactocentric $(X,Y,Z)$ to bar
aligned coordinates $(x,y,z)$ through a rotation in the $(X,Y)$ plane defined as:
\begin{gather}
    \label{eq:bar_coords}
    \begin{pmatrix} x\\ y\\ z \end{pmatrix} 
    = 
    \begin{pmatrix}
    \cos \theta & -\sin \theta  & 0 \\ 
    \sin \theta & \cos \theta  & 0  \\ 
    0 & 0  & 1 
    \end{pmatrix} 
    \begin{pmatrix} X\\  Y\\  Z \end{pmatrix}
\end{gather}
where the angle $\theta$ is defined clockwise about the
$Z$ axis from our line of sight to the Galactic Centre ($X$ axis). Our
model then has 5 parameters denoted as $\mathbf{\Theta} =
(R_{s},Z_{s},x_{0},y_{0},z_{0},\beta,\theta)$, where we have introduced
$\beta$ to represent the central density fraction of bar to disc:
$\rho_{0}^{\textup{bar}} / \rho_{0}^{\textup{disk}}$. 
We will hereafter refer to this as model 1 (M1). For the purposes of generality, we also consider a disk model of the form:
\begin{equation}
    \rho_{\textup{disk}}(R,Z) = \rho_{0}^{\textup{disk}}\,   \textup{exp}\,\left (  -\sqrt{(R^{2}+R_{0}^{2})/R^{2}_{s}} \right ) \textup{exp}\,\left (  -|Z|/Z_{s} \right ) 
    \label{eq:disc_density2}
\end{equation}
now including a softening parameter $R_{0}$. The combination of this softened exponential disk model with the boxy bulge model above will be referred to as model 2 (M2). 

\begin{figure}
\centering
	\includegraphics[width=\columnwidth]{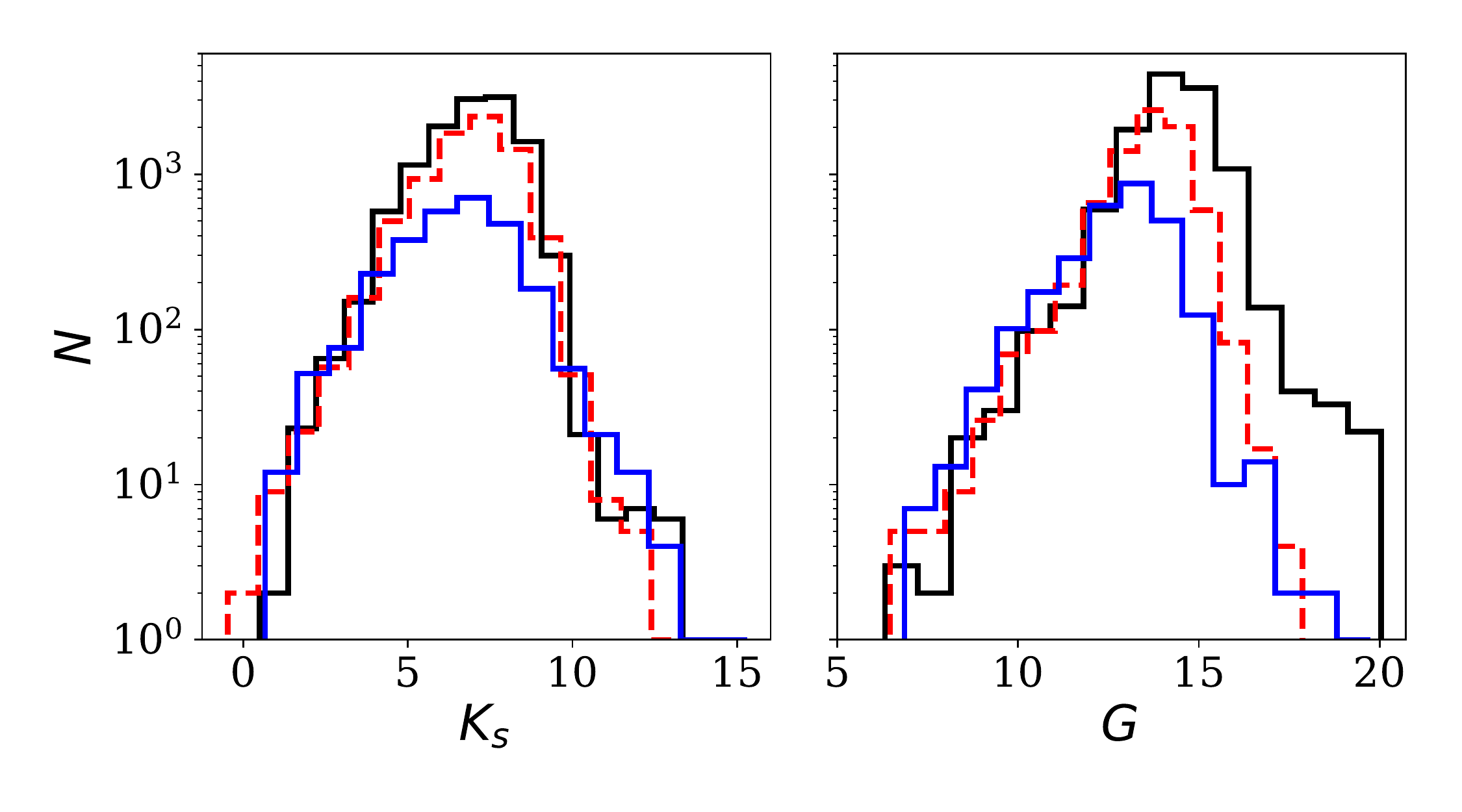}
    \caption[width=\columnwidth]{Distributions of apparent magnitude for our Mira sample, without any extinction corrections applied, in both the 2MASS $K_{s}$ and $Gaia$ $G$ photometric bands. Black solid lines show the Miras lying in the latitude range $0^\circ < |b| < 5^\circ$. The high extinction in this region is visible as an excess of stars at faint magnitudes. The red dashed lines correspond to those lying at $5^\circ < |b| < 10^\circ$ and blue solid lines at $10^\circ < |b| < 20^\circ$.}
\label{fig:LF}
\end{figure}

To transform from heliocentric coordinates $(D,\ell,b)$ to
Galactocentric cylindrical polar $(R,\phi, Z)$ or bulge-aligned Cartesian coordinates $(x,y,z)$, we require the Jacobians:
\begin{equation}
        \label{eq:Jac1}
        \frac{\partial \left ( R,Z,\phi \right )}{\partial \left ( D,\ell,b \right )} = \frac{D^{2}\cos b}{R},\qquad\qquad
        \frac{\partial \left ( x,y,z \right )}{\partial \left ( D,\ell,b \right )}
          = D^{2}\cos b
\end{equation}
We must also account for the spatial footprint over which we can
efficiently observe Miras. Given that we are adopting photometric
distance estimates, this will be a function of the survey
photometry. In Fig.~\ref{fig:LF}, we show the apparent magnitude
distribution of our Miras in both the 2MASS $K_{s}$ and $Gaia$ $G$
bands for different Galactic latitude bins. No reddening correction
has been applied for this figure, so we can assess latitude dependence
of the luminosity function. In both the 2MASS $K_{s}$ and $Gaia$ $G$ bands, we see at low latitudes there is an excess of faint sources, those suffering from regions of high extinction. Accordingly, we only consider stars with $|b| > 5^{\circ}$ in our analysis, \textcolor{black}{removing highly extincted sources together with the region in which the extinction maps of \citet{SFD_map} become highly unreliable. We make use of the \citet{Gonzalez_2012} IR bulge extinction map, which covers the VVV footprint, to further assess the extinction effects of our sample. We find that $\sim  1\%$ of our Miras have $A_{K_{s}}$ values greater than 0.15 at latitudes beyond $5^{\circ}$.  } 

We further restrict our analysis to stars bound by $K_{s}^{max} = 11$ to mitigate any source detection deficiencies at faint magnitudes. We also see the luminosity functions truncate at bright magnitudes, as sources saturate, and thus we define a lower limiting magnitude for our sample at $K_{s}^{\textup{min}} = 4.5$.
The volume over which we compute our likelihood in
eqn~(\ref{eq:lnL}) is bound by these imposed magnitude limits, with corresponding distance limits obtained from Eq.~\ref{eq:D}. Further, though the contribution of high latitude halo stars is small, as is evident from Fig.~\ref{fig:lb}, we also restrict our analysis to
stars lying at $|b| < 20^{\circ}$. No Galactic
longitudinal restriction is made in this analysis. Thus, our
selection function is simply:
\begin{equation}
  S\left ( D,\ell,b \right ) = \begin{cases}
    1 & {\rm if }\ D \in \left ( D_{\textup{min}}, D_{\textup{max}} \right )\  
    \textup{and } b \in \left ( b_{\textup{min}}, b_{\textup{max}} \right ), \\ 
    0 & \text{else}.\\
    \end{cases}
\end{equation}
The bounding volume is a function of Mira luminosity,
which itself depends on pulsation period. Thus, the variation of volume as a function of Mira period must be accounted for in our likelihood calculation, achieved by splitting our likelihood into sub-period bins.  
%
%
%
A further important consideration is the uncertainty in the distance
assigned to each Mira, stemming from the width of the PL sequence in
Fig.~\ref{fig:lmc}. We neglect errors on the periods assigned to the
Miras by $Gaia$ as they are typically less than 1$\%$ for our
sample. We model the distance modulus distribution of our LMC Miras as a Gaussian and fit using least squares regression, obtaining a standard deviation of $\sigma = 0.23$. This distribution width is then convolved into our likelihood computation. \textcolor{black}{We note that the distance uncertainty accounted for in this convolution outweighs that required for sources with $A_{K_{s}} \sim 0.15$. As mentioned earlier in the text, a negligible fraction of our Miras bear values larger than this, based on the extinction map of \citet{Gonzalez_2012}.}

We sample our posterior distribution in Markov Chain Monte Carlo (MCMC) fashion, applying flat priors, to make an inference on our model parameters. To do so, we implement the \texttt{emcee} sampler provided by \citet{emcee}. Numerical computation of the normalising integral is expensive and so we initially do so over a regular grid in our model parameter space. We then linearly interpolate over the grid, enabling posterior samples to be drawn for each MCMC step in parameter space. We choose to model the four period bins shown in Fig.~\ref{fig:XY_RZ} separately. Applying the selection criteria outlined in this section, we retrieve 1,198 Miras to model with periods in the range of $100-200$ days. For the two intermediate period bins, we obtain 1,175 and 2,942 Miras respectively with 3,155 stars in our high period bin. Hereon we will reference these bins as P1, P2, P3 and P4 in order of increasing period.

\textcolor{black}{Using the period age relations of \citet{Feast_2009} and \citet{Wyatt_1983}, we estimate stars belonging to P1 bear ages of $\sim 10$ Gyr and older. For stars in P2 and P3, we reckon ages to lie from $\sim 10$ Gyr down to $\sim 7$ Gyr. For the long period Miras in P4, we estimate their ages to lie in the range $\sim 5-7$ Gyr. Given that we have no metallicity information for the Miras in our sample, we probed its effect on age estimation using the period-metallicity relation of \citet{Feast_2000}. Drawing mock metallicity values from a Gaussian with a standard deviation of $0.3$ dex (as judged from \citet{Hayden_2015} and \citet{Gonzalez_2015}), we estimate this to impact the age estimates by $\sim 1$ Gyr. We acknowledge the inherent uncertainty in these age estimates, but assign them to provide a sense of the age distribution of our Mira sample. Irrespective of metallicity uncertainties, it remains true that the development from short period to long period tracks Miras from old and metal poorer to young and metal richer. We use the nomenclature "old" and "young" purely in the context of our sample.}

\textcolor{black}{It has been observed by \citet{Kordopatis_2013} that at metallicities less than $1.5$, halo stars can dominate over disk stars in the solar neighbourhood. Recently, \citet{Arentsen_2020} posit this may also be the case in the inner galaxy by analysing the kinematics of metal poor bulge stars. Stars of such low metallicity are extremely rare however and very few M giants have been observed with metallicities broaching this regime \citep[see e.g.][]{Rich_2005, Sch_2016}. We therefore expect such effects to be negligible in our sample, even for our lowest period Miras in P1.}

%
%
%

\subsection{Simulated data}

We first implement our sampling method on a mock data set. We generate
a population of stars comprised of a disk and a bar component. For the
disk population, we draw cylindrical coordinates $(R,Z)$ from
exponential distributions with parameters $R_{s} = 3.5$ kpc and $Z_{s} =
1.0$ kpc to mimic our double exponential model of equation~\ref{eq:disc_density}.
\begin{figure}
\centering
	\includegraphics[width=\columnwidth]{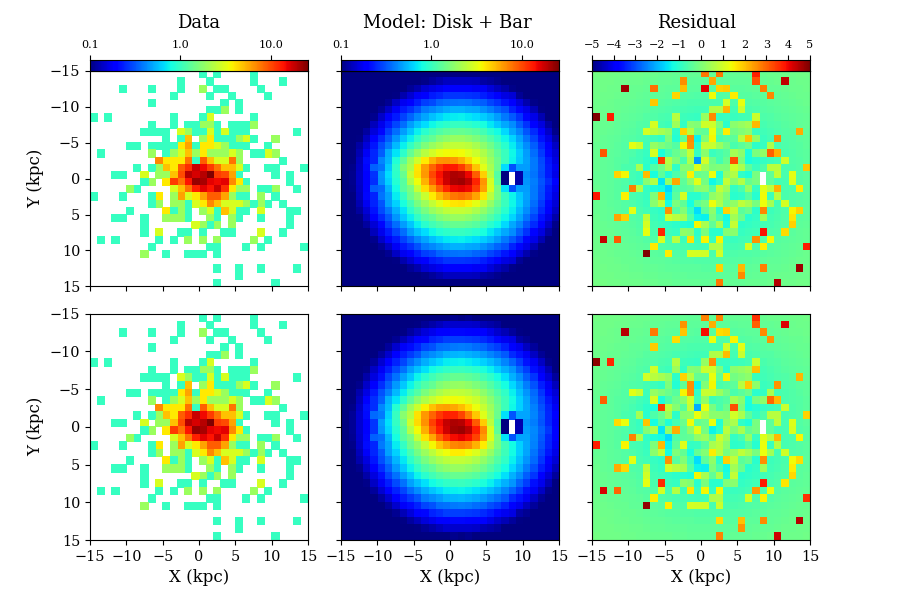}
    \caption[width=\columnwidth]{Comparisons of our simulated data with two models: the initial simulated model with parameters $\mathbf{\Theta}_{\textup{sim}}$ and that recovered from our MCMC inference procedure. The left columns shows the stellar number density projected into the Galactocentric $(X,Y)$ plane for the models and the middle column for the simulated data. \textcolor{black}{The lack of stars around the Solar position in the model panels stems from the limiting magnitude mask imposed when modelling the stellar density.} The right columns shows the residual, scaled by the Poisson error in each corresponding model pixel.}
\label{fig:sim_xy_residual}
\end{figure}
Similarly for the bar, we draw $(x,y,z)$ coordinates from the triaxial Gaussian distribution in eqn~(\ref{eq:bar_density}) fixing the parameters at $(x_{0},y_{0},z_{0}) = (2.0,1.3,0.7)$ in units of kpc. We prescribe the stars period values ranging from $100-200$ days, drawn from a narrow Gaussian centered on $P = 150$ days with a width of 50 days. Further, we fix the ratio of bar stars to disk stars to approximate that of our real data set in this period range, providing a simulated value for $\beta$ as 2.04.

Following the methods of Section~\ref{sec:model}, we infer the density profile parameters of our simulated data and make a comparison to those that we initially implemented in the generation of the mock data. We do so in Fig~\ref{fig:sim_xy_residual} where we show the two sets of residuals in the $X-Y$ plane; the top row being that of simulated model and the bottom row using the model parameters recovered from our inference procedure. The figure demonstrates excellent agreement between the two rows with negligible difference between the two residuals. 
\begin{table}
\centering
\begin{tabular}{ |c|c|c| } 
\hline
 & Input & Recovered  \\
 \hline 
 $R_{s}$ (kpc) &3.5 & $3.61^{+0.14}_{-0.16}$  \vspace{1mm} \\ \vspace{1mm}
 $Z_{s}$ (kpc) & 1.0&$1.10^{+0.05}_{-0.05}$ \vspace{1mm} \\ \vspace{1mm}
 $x_{0}$ (kpc) &2.0 &$2.05^{+0.14}_{-0.15}$ \vspace{1mm} \\ \vspace{1mm}
 $y_{0}$ (kpc) & 1.0&$0.95^{+0.08}_{-0.07}$ \vspace{1mm} \\ \vspace{1mm}
 $z_{0}$ (kpc) & 0.7& $0.73^{+0.05}_{-0.04}$ \vspace{1mm} \\ \vspace{1mm}
 $\beta$ & 2.04 & $2.12^{+0.47}_{-0.39}$ \vspace{1mm} \\ \vspace{1mm}
 $\theta$ (deg) & 20 & $17.48^{+3.7}_{-3.5}$ \\ 
 \hline
\end{tabular}
\caption{Table shows the simulated input parameters used to generate our mock data and the recovered parameters from our fitting procedure.}
\label{table:sim}
\end{table}
\section{Results}
\begin{figure}
\centering
	\includegraphics[width=\columnwidth]{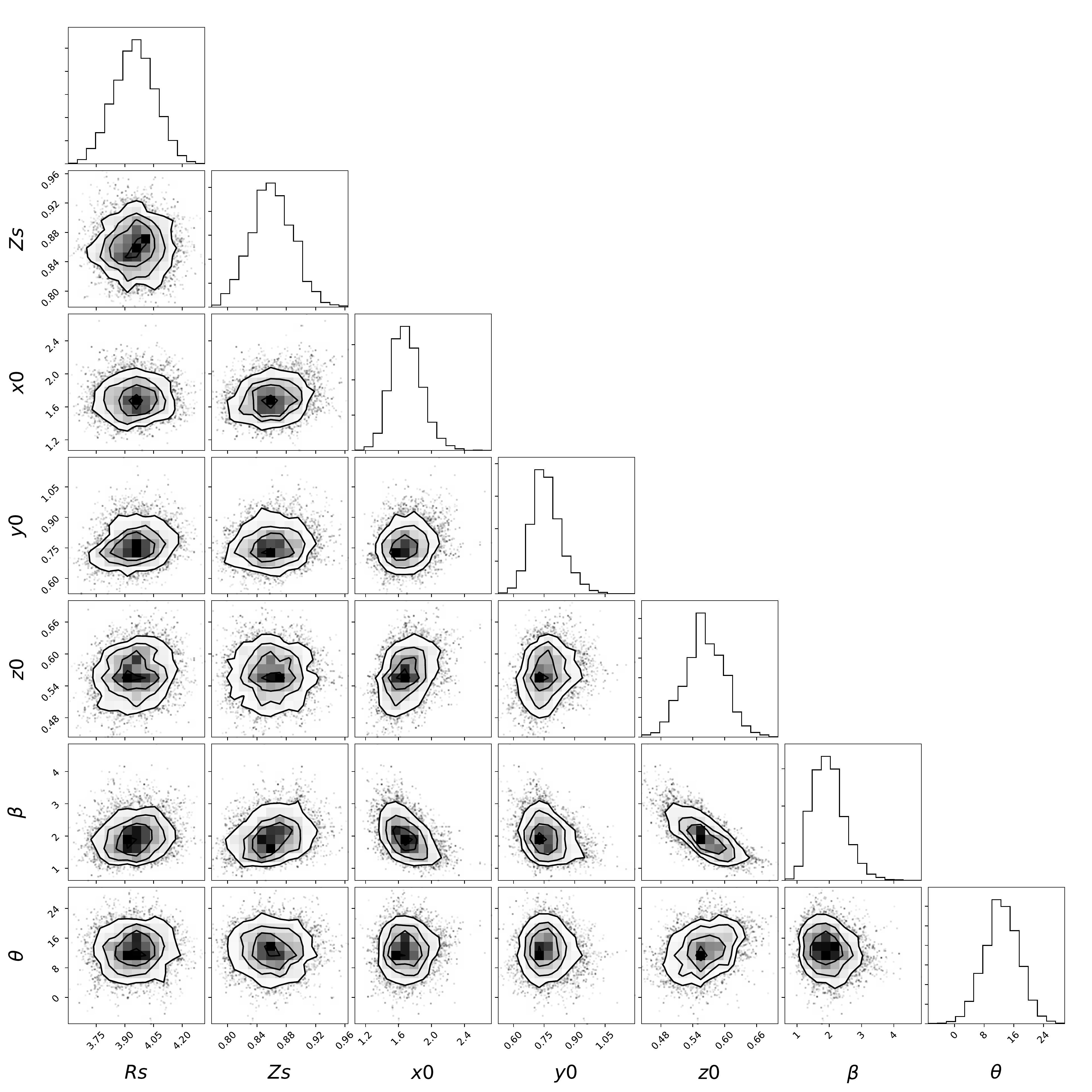}
    \caption[width=\textwidth]{Marginalised posterior distributions for the model parameters of the Mira population in the period range 200-250 days recovered from our MCMC fit.}
\label{fig:corner}
\end{figure}

\begin{table*}
\begin{tabular}{ cccccccc } 
\hline
\vtop{\hbox{\strut Period bin}} &
\vtop{\hbox{\strut \, $R_{s}$}\hbox{\strut(kpc)}} &
\vtop{\hbox{\strut \, $Z_{s}$}\hbox{\strut(kpc)}}&
\vtop{\hbox{\strut \, $x_{0}$}\hbox{\strut(kpc)}}& 
\vtop{\hbox{\strut \, $y_{0}$}\hbox{\strut(kpc)}}& 
\vtop{\hbox{\strut \, $z_{0}$}\hbox{\strut(kpc)}}& 
$\beta$ & 
\vtop{\hbox{\strut \, \, $\theta$}\hbox{\strut(deg)}}
\\ \hline

P1 & $3.78^{+0.13}_{-0.13}$ & $1.01^{+0.05}_{-0.04}$ & $1.85^{+0.23}_{-0.15}$ & $0.81^{+0.07}_{-0.06}$ & $0.64^{+0.04}_{-0.03}$ & $3.29^{+0.65}_{-0.55}$ & $1.27^{+3.20}_{-2.80}$   
\vspace{1mm} \\ \vspace{1mm} 

P2 & $3.96^{+0.10}_{-0.11}$ & $0.86^{+0.03}_{-0.03}$ & $1.69^{+0.19}_{-0.17}$ & $0.76^{+0.08}_{-0.06}$ & $0.56^{+0.04}_{-0.04}$ & $1.95^{+0.51}_{-0.47}$ & $12.71^{+4.68}_{-4.60}$
\vspace{1mm} \\ \vspace{1mm}

P3 & $4.47^{+0.11}_{-0.09}$ & $0.66^{+0.02}_{-0.01}$ & $1.68^{+0.12}_{-0.11}$ & $0.73^{+0.05}_{-0.05}$ & $0.55^{+0.02}_{-0.03}$ & $1.59^{+0.36}_{-0.26}$ & $21.16^{+3.27}_{-3.41}$
\vspace{1mm} \\ \vspace{1mm}

P4  &  $4.52^{+0.10}_{-0.10}$ & $0.51^{+0.01}_{-0.01}$ & $2.02^{+0.32}_{-0.28}$ & $0.56^{+0.07}_{-0.07}$ & $0.55^{+0.04}_{-0.04}$ & $0.54^{+0.17}_{-0.11}$ & $21.08^{+2.28}_{-2.66}$
\\
\hline
\end{tabular}
\caption{We list the recovered parameters of our model, for each Mira period bin, following the procedure detailed in Section~\ref{sec:model}. \textcolor{black}{For the old stars (P1), they have age estimates of $\sim 10$ Gyr. Stars in P2  and P3 bear ages of $7-10$ Gyr. For the youngest Miras, we estimate their ages to be $\sim 5-7$ Gyr. We base these value on the Mira age estimates of \citet{Wyatt_1983} and \citet{Feast_2009}.}}
\label{table:mcmc}
\end{table*}
We apply our parameter fitting procedure to each of the three sub-populations of Miras shown in Fig.~\ref{fig:XY_RZ}, probing the evolution of disk and bar morphology with pulsation period. Fig.~\ref{fig:corner} demonstrates the resultant posterior distributions obtained from our fits, with the figure being representative of the intermediate period population P2 and we list the full set of parameters for each of the period bins in Table~\ref{table:mcmc}. We show the full evolution of our model parameters in each period bin in Fig.~\ref{fig:pars} with our main model M1 identified by red markers.

\subsection{Miras in the Disk}

\begin{figure*}
\centering
	\includegraphics[width=\textwidth]{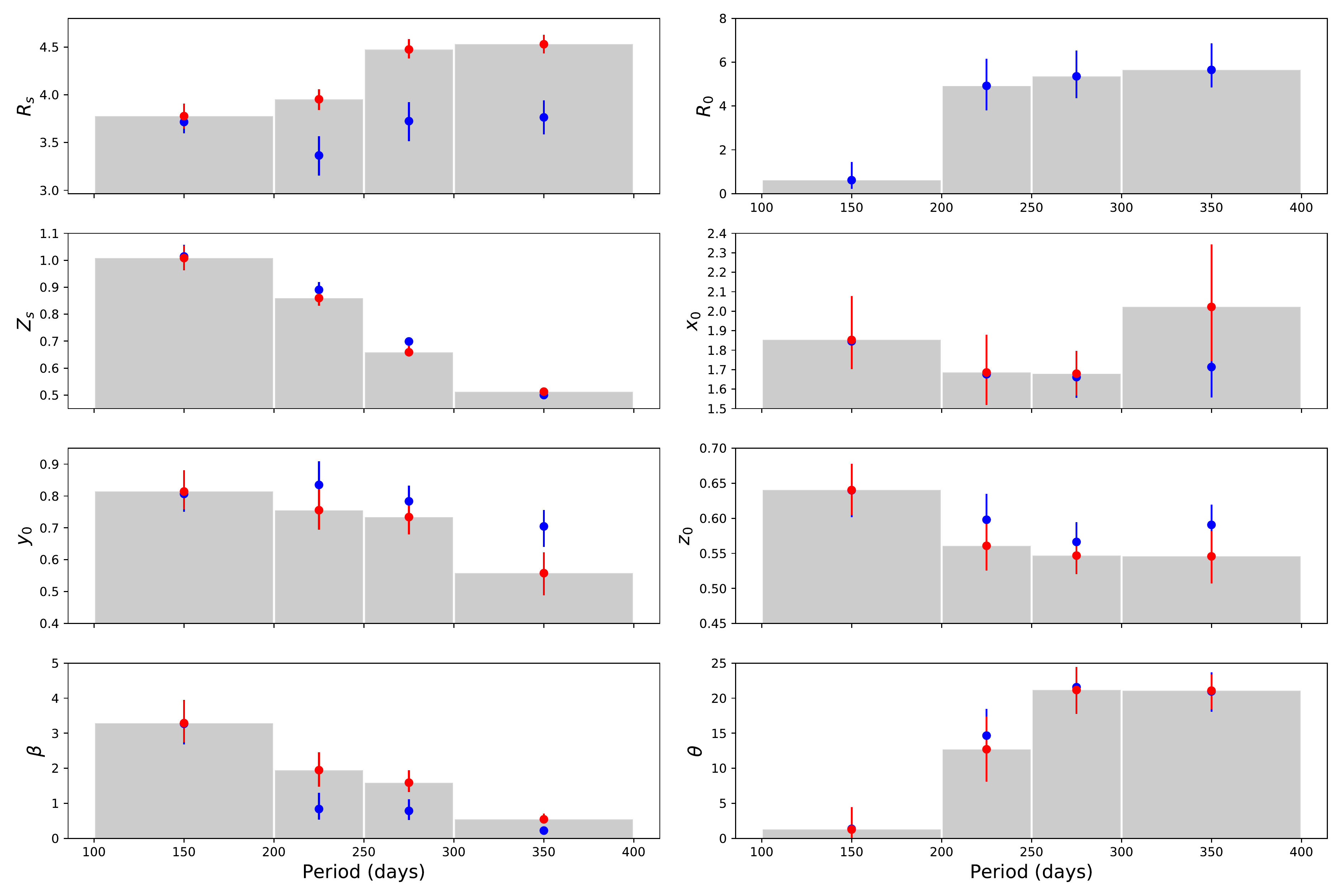}
    \caption[width=\textwidth]{We show the evolution of recovered model parameters as a function of Mira pulsation period. Red scatter points correspond to the values given in Table.~\ref{table:mcmc}, being the parameters of our main model, M1. The parameters show strong evolution as the older Miras make up a more radially concentrated and thicker disk, evolving toward a thinner and more extensive disk for younger stars. The bulge/bar length is consistent with being constant across all bins, but appears to narrow, becoming thinner for the youngest of Miras. The bar angle shows strong progression from being very nearly aligned with the Solar line of view at old ages up to $\theta \sim 21^{\circ}$ for the younger Miras. Blue scatter points correspond to our softened exponential disk M2, and largely show the same distribution as those of M1. This differs in the case of the radial scale length of the disk, for which we see a largely constant value across all period bins. This is compounded with the progression of the $R_{0}$ parameter evolving from near zero to $5-6$ kpc. The grey bars indicate the width of each period bin with the scatter points located at the central value.}
\label{fig:pars}
\end{figure*}

From Fig.~\ref{fig:pars}, the disk parameters of our main model (M1) show an evolution in agreement with the qualitative picture obtained from Fig.~\ref{fig:XY_RZ}; as we progress from a high to low period population of stars, the thickness of the disk increases reaching a scale height of $\sim 1$ kpc, comparable to traditional estimates for the thick disk. Similarly, we see an evolution in the radial scale length of the disk, originating at low values for low periods and growing on increasing period. Such a smooth evolution of the disk's stellar profile complements the findings of \citet{Bovy_2012}, who show there is a continuous change of the disk scale parameters as a function of stellar position in the $\left [ \alpha/\textup{Fe} \right ]$ versus $\left [ \textup{Fe}/\textup{H} \right ]$ plane. Further, using a sample of $\sim$ 70,000 red giants, \citet{Hayden_2015} mapped out the stellar distribution in the $\left [ \alpha/\textup{Fe} \right ]$ versus $\left [ \textup{Fe}/\textup{H} \right ]$ plane across the Galaxy, spanning a radial range of $3-15$ kpc in the plane and up to $2$ kpc vertically away from the plane. They observed a progression in which $\alpha$-rich stars dominate in the inner disk at large heights but $\alpha$-poor stars dominate further out and closer to the disk plane. This is consistent with the distribution of our disk parameters. We may physically interpret our results to demonstrate the changing morphology of the Galactic disk as a function of chemistry and stellar age, given that the pulsation period of Miras acts as a proxy for both. The older, metal poor stellar population resides in a stubby disk; it is radially restricted but bloated high above the disk plane. As the age of the stellar population decreases and becomes metal enhanced, the stars reside closer to the Galactic plane but stretch out radially. This is a clear signal of inside-out, upside-down formation of the Milky Way disk over a stellar age spanning $\sim 5-10$ Gyr. The study of Milky Way-like galaxies in the Eris simulations catalogue by \citet{Bird_2013} found the oldest of stellar populations to reside in centrally concentrated, puffy configurations with younger cohorts continuously populating radially more extended and vertically thinner disks. Tracing the star formation history of these sub-populations backward, they saw the older stars to have formed during active merging phases in the galaxy's history, dynamically heating the stellar orbits. The larger scale heights of the older stars were seen to be largely inherited from the turbulence associated with the early star-forming disk, imprinted by the high supernova feedback rate and larger rate of dynamical heating by mergers and halo substructure. Younger stars are born out of cooler, more rotationally supported gas at calmer periods in the galaxy's lifetime, and the disk grows radially.
\begin{figure*}
\centering
	\includegraphics[width=\textwidth]{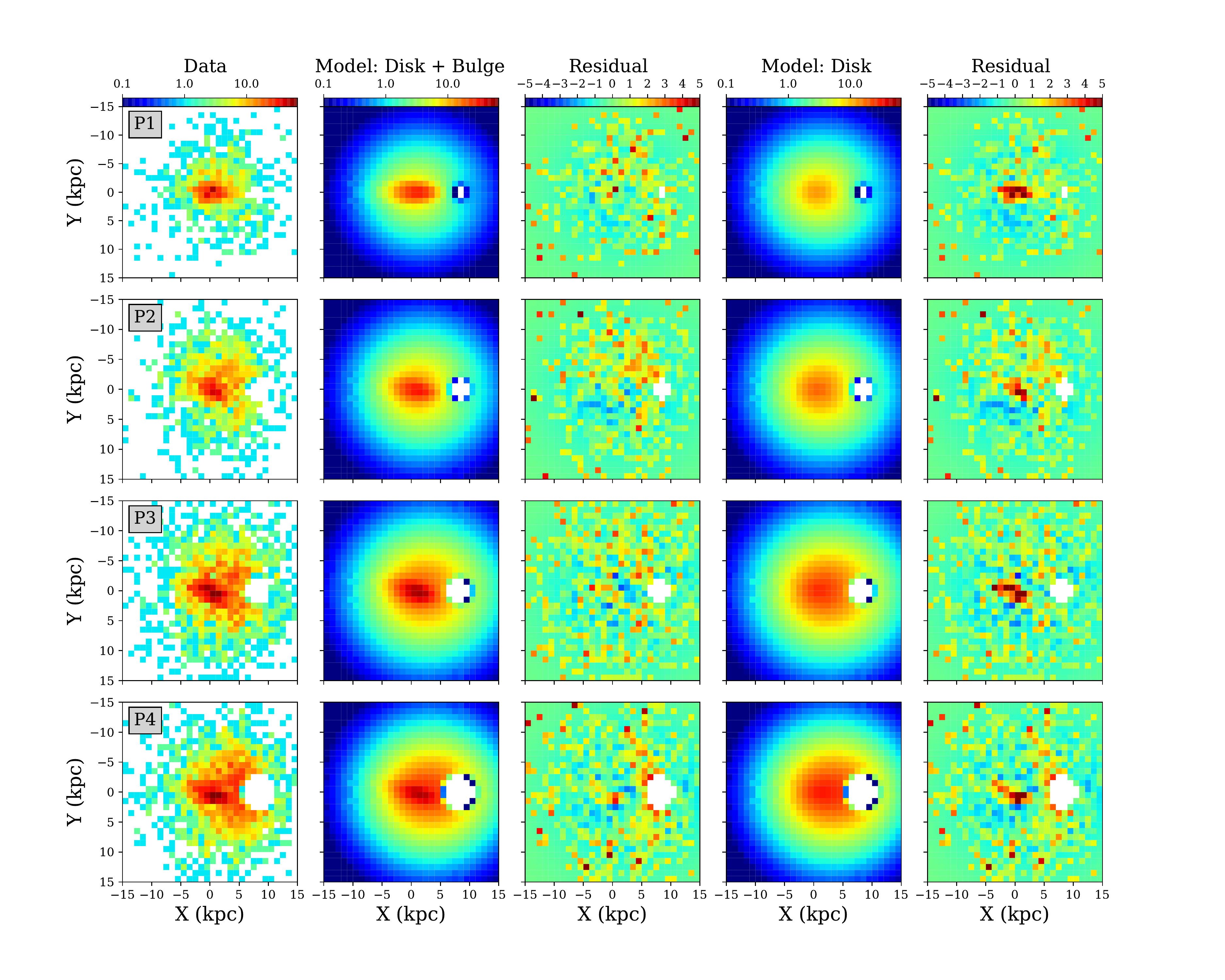}
    \caption[width=\textwidth]{Comparison of our Mira sample modelled in each period bin with our recovered model from the MCMC fit in Galactocentric $(X,Y)$ projection. The Solar position is at (8.3,0) with the $Y$ axis aligned with the direction of Galactic rotation. Top to bottom panels run in order of low period to high period corresponding to stellar ages beyond $10$ Gyr down to $\sim 5$ Gyr. The data column shows the logarithmic number density of the Miras and the model columns show the corresponding predicted number density based on the parameters in Table.~\ref{table:mcmc}. The residuals are computed as the difference in data count and model count, weighted by the Poisson noise in each model. We show two instances of our model: that with an exponential disk + boxy bulge and that with an exponential disk only. The latter highlights the presence of the bulge Miras, made apparent by the strong central residuals in the right hand column.}
\label{fig:XY_residuals}
\end{figure*}
\begin{figure*}
\centering
	\includegraphics[width=\textwidth]{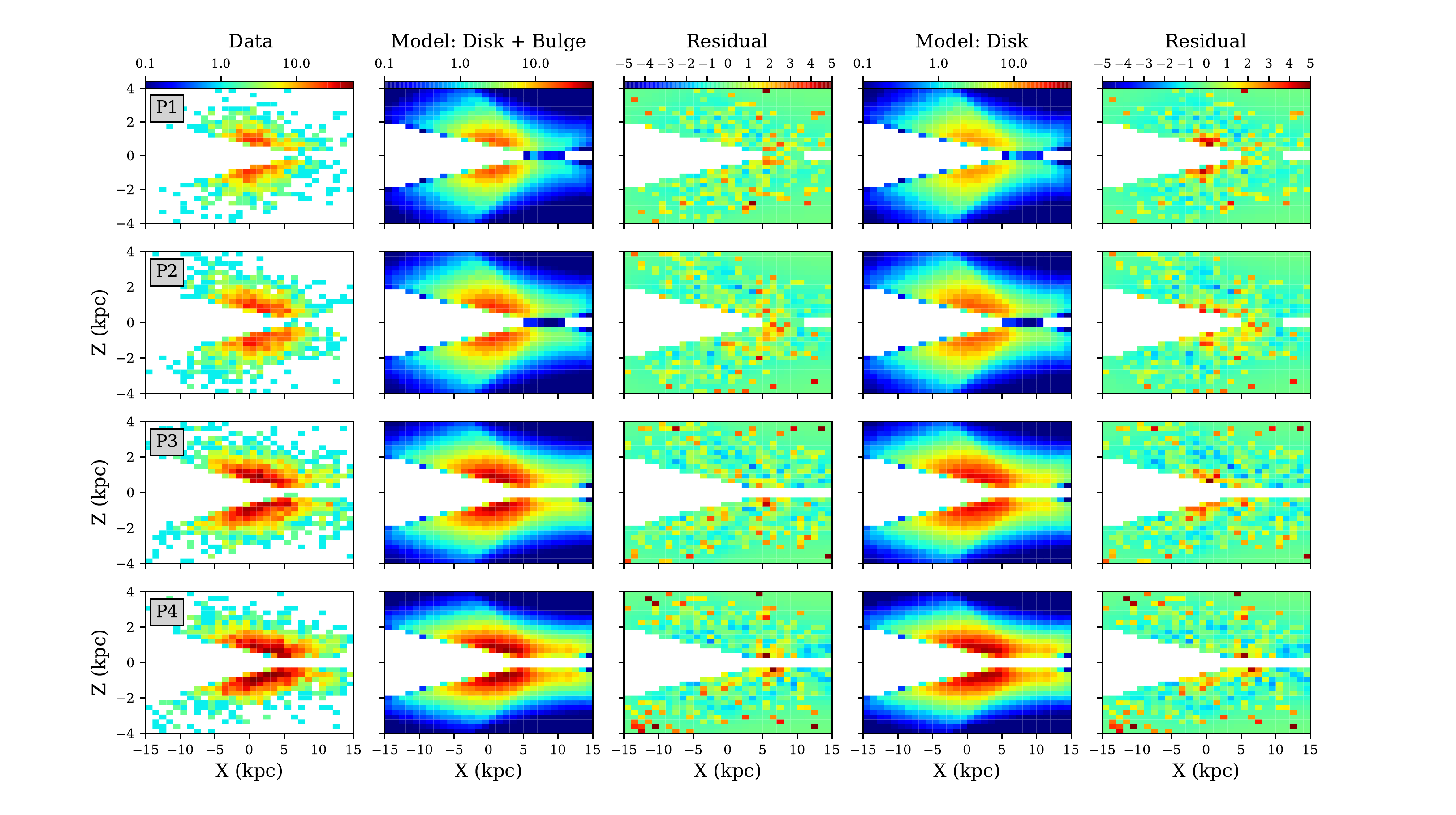}
    \caption[width=\textwidth]{\textit{Top}: Galactocentric $X-Z$ projections of our low period Mira sample. From left to right we show the recovered model from our fit, the data number density and the residuals scaled by the Poisson noise in each model pixel. \textit{Middle} and \textit{Bottom} show the equivalent for the intermediate and high period Mira populations respectively.}
\label{fig:XZ_residuals}
\end{figure*}
\begin{figure*}
\centering
	\includegraphics[width=\textwidth]{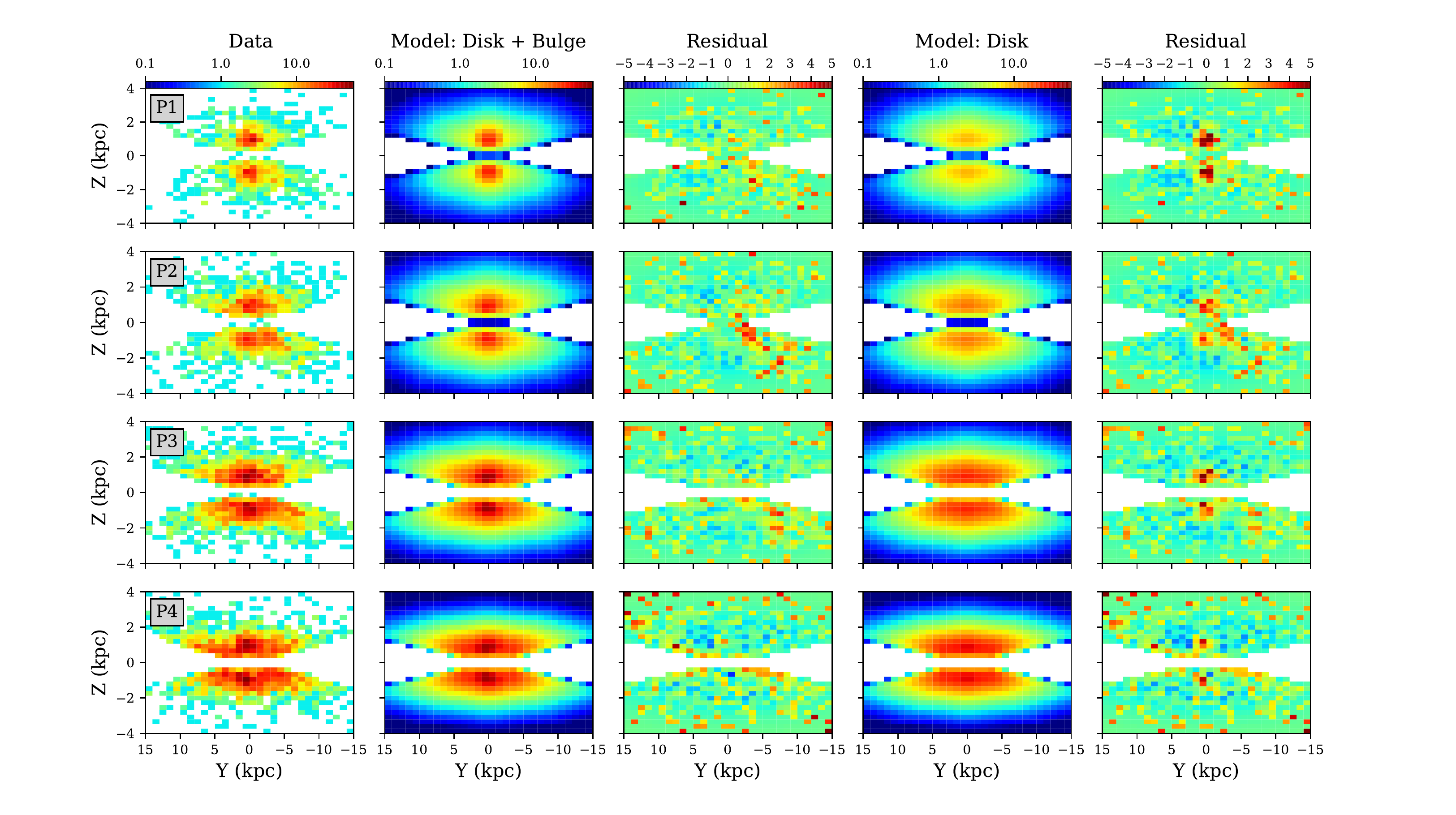}
    \caption[width=\textwidth]{\textit{Top}: Galactocentric $(Y,Z)$ projections of our low period Mira sample. From left to right we show the recovered model from our fit, the data number density and the residuals scaled by the Poisson noise in each model pixel. \textit{Middle} and \textit{Bottom} show the equivalent for the intermediate and high period Mira populations respectively.}
\label{fig:YZ_residuals}
\end{figure*}
Looking at the Galactic $(X,Y)$ residuals of our model M1 in Fig.~\ref{fig:XY_residuals}, we see there exists a strong ring like over-density amongst the youngest of the Miras (P4). The ring coincides approximately with the Solar radius and stretches out to $Y \sim \pm 10$ kpc, being more prominent away from the direction of Galactic rotation (i.e negative $Y$). Galactic rings are a well known phenomenon observed in many external galaxies largely thought to be formed through secular evolutionary processes within the galaxy. In barred galaxies, the bar itself can drive gas radially until it is halted at a resonance. In the case of outer rings, this occurs at the Outer Lindblad Resonance (OLR), with subsequent star formation yielding a stellar ring. For an approximately flat Galactic rotation curve, the OLR should reside at $R_{\textup{OLR}} \sim 1.7R_{\textup{CR}}$ where $R_{\textup{CR}}$ is the co-rotation radius. This is believed to lie in the range of $\sim 3-6$ kpc \citep[see e.g.][]{Gerhard_2011}, with a recent value of $5.7 \pm 0.4$ kpc obtained by \cite{Sanders_2018} using a combined $Gaia$ DR2 + VVV data set. Thus, the OLR is expected to reside somewhere between $6$ kpc and $10$ kpc, in the immediate neighbourhood of the Solar position. Given that the ring-like residual seen in the P4 Miras coincides with this radial range, it is plausible for it to be associated with trapping of stellar orbits around the OLR. Further, it can be gleaned from Figs.~\ref{fig:XZ_residuals} and ~\ref{fig:YZ_residuals} that there are Miras lying beyond $2$ kpc from the disk, manifesting as the fluffy residuals at these heights. We further see a long spur-like structure, constituted of Miras with periods in the range $200-250$ days, in the $(Y,Z)$ projection of Fig.~\ref{fig:YZ_residuals}. Its extent is large, stretching out to $Y \sim -5$ kpc and $\sim 3$ kpc below the plane of the disk. With the ever increasing wealth of stellar information provided by modern surveys, it is becoming evident that the Milky Way disk is not in dynamical equilibrium but displays various signatures of recent perturbations. A wave-like north-south asymmetry of stellar counts was observed by \cite{Widrow_2012}, who suggest an externally driven vertical perturbation may be the cause for such disk heating. With the advent of $Gaia$ DR2, clear evidence of vertical phase space mixing was seen by \cite{Antoja_2018} within the Solar neighbourhood. The perpetrator of such a signature is still debated, with possibilities being a Sagittarius-like object plunging through the disk \citep[see e.g.][]{Binney_2018,Li_2019,Laporte_2019} or even vertical oscillations induced by the buckling of the bar \citep[see e.g.][]{Khoperskov_2019}. Such heating events inevitably kick stars out of the disk to large heights and indeed stellar debris as high as $\sim 10$ kpc may have originated in the disk \citep{Xu_2015,PW_2015}. As the Miras represent a thick disk stellar population, and thus 
in-plane perturbations due to spiral arms and giant molecular clouds etc are likely ineffective, it is reasonable to interpret the origin of the high lying Miras as that of external disk heating, kicking the stars to such heights. 

For our softened exponential model (M2), the general evolution of the parameters is equivalent, except in the case of the radial scale length which is consistent with being constant at all periods. The softening parameter $R_{0}$ shows a strong jump from near zero in P1, consistent with a simple double exponential disk, to $\sim 5-6$ kpc at higher periods. The interpretation for this is not clear and it may simply be that the softened model is acting to smooth out the stellar radial distribution at these periods. For consistency, we show the Galactocentric residuals of this model in the Appendix, but there is little difference to those of M1. 

\textcolor{black}{We note the presence of a model over-density at $(X,Y) \sim (-5,4)$ kpc in our P2 residual panels. This likely stems from combined effects of the $Gaia$ scanning law and sparse light curve sampling, inhibiting period recovery at periods of $\sim 200$ days \citep[see][]{Gaia_LPVs}. Checking the spatial distribution for each of our period bins, as in Fig.~\ref{fig:lb}, we indeed see regions of low detection efficiency in P2, but largely at latitudes less than $5^{\circ}$, a region not included in our modelling.}
%
%
%
%
%
%
\subsection{Miras in the Bulge}

Turning our attention to the bulge, we observe several interesting features in our recovered parameters of M1. Across all period bins, the bulge/bar length shows no clear evolution within the error, with a value of $x_{0} \sim 1.7-1.9$ kpc observed. Further, as the stellar age of the bulge decreases (on increasing period), it appears to become narrower and thinner. Based on the simulations of \citet{Debattista_2017}, this is an expected trend, one that they label 'kinematic fractionation'. They show that through both purely N-body and high resolution, gas dynamical simulations, the resultant bulge morphology correlates strongly with the age of the stellar population in question, seeded by their initial in-plane random motions. The older, hotter stellar populations form a vertically thicker boxy bulge and the younger, cooler stars result in a thinner, peanut-like bulge (see Fig.~15 of \citet{Debattista_2017}). This change in bulge morphology is true for the Miras. The bulge residuals when we model only for the disk in Fig.~\ref{fig:YZ_residuals} clearly show the oldest Miras to extend vertically higher (i.e. thicker) than the younger Miras and our recovered model parameters in Fig.~\ref{fig:pars} clearly show the trend of decreasing width (i.e. become narrower) on increasing period. We note that whilst from Fig.~\ref{fig:XY_RZ} it appears that the bulge/bar length grows for the highest two period bins, our modelling is limited to $|b| > 5^{\circ}$ and so we are insensitive to this in plane bulge growth. Our latitudinal mask also inhibits the vertical extent over which we model the data (e.g. see Fig.~\ref{fig:XZ_residuals}) and this may be a reason why we do not observe a continuous decline in the bulge scale height.

For the oldest Miras, the scale length ratios of the bulge/bar are $[1:0.44:0.35]$ in the order of $[x_{0}:y_{0}:z_{0}]$. For the two intermediate populations, we obtain $[1:0.45:0.33]$ and $[1:0.43:0.33]$ with the youngest of Miras showing $[1:0.28:0.27]$. These values are comparable to the three dimensional structure of the Galactic bulge found by \citet{Pietrukowicz_2015} in their study of OGLE-IV RR Lyrae, as well as the that of the RC bulge stars studied by \citet{Bissantz_2002} and \citet{Rattenbury_2007}. We find the elongation of the bar to be slightly less than that found by \citet{Wegg_2013}, who utilised RC stars in the VVV survey to derive axis ratios of $[1:0.63:0.26]$ \citep[see][for discussion]{Zoccalli_2019}. This discrepancy may stem from the fact that they probe the bar region to lower latitudes ($|b| < 5^{\circ}$) than accessible to us here, though the extensive modelling efforts of VVV RC stars by \citet{Simion_2017} yielded a best fitting bulge axis ratio of $[1:0.44:0.31]$. Further, \citet{Wegg_2015} analysed a sample of RC stars residing outside of the bulge, highlighting the existence of a vertically 'thin' and 'super-thin' component associated with increasingly younger stellar populations, down to $\sim$ 1 Gyr. They also see a smooth transition between the boxy bulge and long bar through the continuous decline in the scale heights of their fits.  This is corroborated by \citet{Cabrera_lavers_2007,Cabrera_lavers_2008} who find two triaxial structures residing in the inner galaxy: a thick boxy-peanut bulge and a thinner long stellar bar. The claim of a long bar is not novel with \citet{Gonzalez_2012,Amores_2013,Lopez_2007,Hammersley_2007} all finding such a structure, located at a position angle of $\sim 40-45^{\circ}$, significantly offset from that typical of the main bar, inspiring the notion of two separate bulge/bar structures within the galaxy. However, the simulations of \cite{Martinez_Valpuesta_2011} and \cite{Romero_Gomez_2011} imply that such an offset may be a result of volume projection effects in the stellar counts and possible asymmetries at the end of the bar, which develop leading ends induced by interactions with adjacent spiral arm heads. Such an effect appears to have been observed in \textcolor{black}{Sloan Digital Sky Survey (SDSS)} imaging of the galaxy MCG+07-28-064 where \cite{Peterken_2019} observe a change in the bar angle at the youngest of stellar ages. They interpret this to be an effect of star formation occurring on the leading edge of the bar, wherein the young stars have not had ample time to fully mix in their orbits through the bar potential. It would therefore seem a single central structure may suffice; a central boxy bulge extending to a longer, thinner in-plane bar which can couple to the spiral arms yielding non-symmetric bar ends. In this context, the evolution of the dimensions of our Mira populated bulge are consistent with known trends of the bulge/bar system in that the width and height of the structure decreases on decreasing stellar age, as observed by \citet{Wegg_2015} and in the extensive simulations of \citet{Debattista_2017}. We note that, whilst less pronounced, the results of M2 are also consistent with this evolutionary picture of bulge/bar structure. 
%
\begin{figure*}
\centering
	\includegraphics[width=\textwidth]{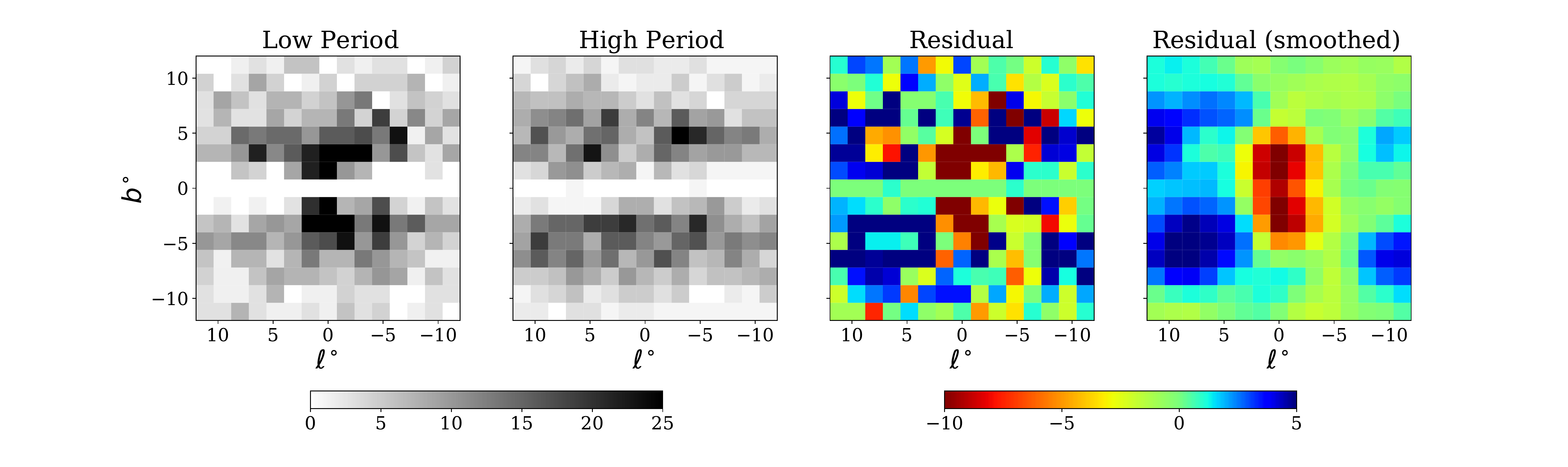}
    \caption[width=\textwidth]{We show the distribution of the longest and shortest period Miras in our sample for the inner $12^{\circ} \times 12^{\circ}$ of the Galaxy in the grey scale panels. The leftmost of these corresponds to the low period Miras whilst the right that of the long period Mira. We further restrict ourselves to stars within a Galactocentric cylindrical radius less than $5$ kpc to reduce foreground contamination. Normalising the stellar number density across the two period bins, we compute the residual between the two. We smooth the image by applying a Gaussian filter of width $1.6^{\circ}$ over the residual. This is displayed in the two rightmost coloured panels. Blue over-densities correspond to regions dominated by the long period (young/metal rich) Miras and red those by the old (old/metal poor). The ancient Miras peak centrally, on the minor axis. Away from this region, the young Miras dominate in two fields either side of the Galactic center, with greater vertical extent on the near side of the bulge. This is indicative of the younger Miras constituting an peanut-shaped bulge inclined to our line of sight.}
\label{fig:Xshape}
\end{figure*}

However, it is striking that the orientation of the bulge Miras evolves on increasing period; the oldest Miras orient themselves along our line of sight and the younger Miras reach an inclination angle of $\theta \sim 21^{\circ}$ as seen in Fig.~\ref{fig:pars}. The OGLE-III bulge RR Lyrae studied by \citet{Dekany_2013} using VVV photometry revealed a slightly elongated stellar structure, with an inclination angle of $\sim 12.5^{\circ}$ amongst a more spheroidal population. Further, the bulge OGLE-IV RR Lyrae sample studied by \citet{Pietrukowicz_2015} resulted in an inclination angle of $\sim 20^{\circ}$. Previous analysis of bulge RC stars \citep[see e.g.][]{Bissantz_2002,Babusiaux_2005,Rattenbury_2007,Simion_2017,Wegg_2013} have constrained the angle of the bulge/bar to lie in the range of $\sim 20^{\circ}-30^{\circ}$. \textcolor{black}{Recent modelling of the bulge by \citet{Coleman_2019} determines the bar angle to lie in the range $18^{\circ} - 32^{\circ}$ through both parametric and non-parametric modelling of VVV RC bulge stars.} The two youngest Mira populations \textcolor{black}{(P3 and P4)} in our study are consistent with these findings and bear ages of $\sim 5-9$ Gyr. From Fig.~\ref{fig:XY_residuals}, the presence of a central bulge/bar-like feature is evident at all periods from the large residuals in the centre of the galaxy when we fit only for an exponential disk. Curiously, these residuals are clearly very nearly aligned with our line of sight towards the Galactic centre for the lowest period population. Our best fit parameters for these old Miras do not suggest they are the most spheroidal of our sample, as we may expect for the oldest population. The extent of the residuals in the $(X,Z)$ and $(Y,Z)$ planes in Figs.~\ref{fig:XZ_residuals} and ~\ref{fig:YZ_residuals} affirm this, with their vertical extent constrained below $2$ kpc from the plane but reaching out to $\sim 2.5$ kpc along the Galactocentric $X$ coordinate. An evolution of bar inclination angle with Mira pulsation period has been observed by \citet{Catchpole_2016} who observe distinct bulge/bar structures occupied by the oldest and youngest of Miras in their sample. The ancient Miras (P1) show no apparent inclined structure, whereas the younger Miras form a tilted bar-like structure. This change of orientation is consistent with the difference we see between our shortest period bin and the longer period Miras. This too was observed by \citet{Debattista_2017}, who found that the oldest stars in their simulations showed very weak coupling to the inclined bulge/bar structure, as well as by \citet{Dekany_2013} in their bulge sample of RR Lyrae. \textcolor{black}{This has been corroborated by \citet{Prudil_2019} in their recent analysis of OGLE-IV RR Lyrae finding no evidence of a bar association}. For the stars in P2, we find an inclination of $\sim 13^{\circ}$, comparable to that of the very central RR Lyrae distribution of \citet{Dekany_2013}. However, from Fig.~\ref{fig:XY_residuals}, it is unclear how strong a bulge signal exists in this period bin. The bulge-like residual in the disk only fit is weak and confined only to the very central region. 
Thus our model may be suffering from the sparseness of the data in this narrow period range when fitting for the bulge/bar parameters, given our latitude restrictions. Alternatively, this period bin may be including Miras from two separate bulge populations, one inclined at $\sim 21^{\circ}$ and another not, resulting in an intermediate bulge/bar angle. Aside from the uncertainty in this period range, the evolution of the bar angle from the oldest of Miras to the youngest is clear from the data where the ancient Miras look to be decoupled from the inclined boxy bulge/bar. It is possible therefore that these Miras are old enough to have been born before the formation of the bar and its subsequent buckling. If they were kinematically hot enough, it is possible they did not efficiently couple to the early bar thus maintaining their distinct structure. The prominent elongation of this structure may be a consequence of our imposed latitude cut or an observational bias due to potential incomplete latitudinal sampling of Miras in the $Gaia$+2MASS cross-match.  

It has been predicted from both N-body and star forming simulations that upon observing younger/metal-richer bulge populations, their morphology should tend toward that of a peanut shape with a stronger X-shape signature compared to the older/metal poorer bulge stars \citep[see e.g.][]{Debattista_2017,Fragkoudi_2018}. Indeed, observations have shown this to be the case with the older, metal poor stars being centrally concentrated and axisymmetric in their spatial distribution. The young and metal rich stars are consistent with belonging to a boxy/peanut shaped bulge bearing the X-shape signature characteristic of such a structure inclined to the line of sight \citep[see e.g.][]{Rojas_2014,Wi16,Zoccali_2017}. Accordingly, we seek to find such a contrast in bulge spatial distribution in our Mira sample. Selecting only the oldest and youngest period bins of our sample, we show their density distribution in Fig.~\ref{fig:Xshape}. On computing the residuals between these period bins, we can identify the regions in which the two sub-populations dominate the overall stellar density. Indeed, we see the old/metal poor Miras to lie centrally and extend vertically along the bulge/bar minor axis, whilst the young/metal rich Miras show four peaks in density, distributed asymmetrically about the Galactic centre. We interpret this as evidence that the young Miras lie in a boxy/peanut configuration with a viewing angle of $\sim 21^{\circ}$ yielding the apparent X-shape. Our Galactic latitude mask at $|b| < 5^{\circ}$ inhibits us from making a direct comparison to our smooth bulge/bar model, as this signal exists primarily at low latitudes in a region of low stellar density. Regardless, the signal we recover is in excellent agreement with the predicted stellar age distribution of \citet{Debattista_2017} (see their Fig.~22) and is reminiscent of the general X-shape morphology uncovered by \citet{Ness_2016}.

\textcolor{black}{Finally, we note that there has been long discussion of long period bulge Miras in the context of the apparent dearth of young stars seen to reside in the central region of the Galaxy \citep[see e.g.][]{ Zoccali_2003,Valenti_2013}. Both \citet{Renzini_2018} and \citet{Clarkson_2011} have set a limit on bulge stars younger than $\sim 5$ Gyr, coincident with our lower age limit. Miras covering a wide range of periods residing in the bulge/bar have been studied by \citet{Whitelock_1991} who consider the possibility that the long period Miras are in fact the progeny of binary mergers rather than a distinctly young population, a notion put forth by \citet{Renzini_1990}. However, the majority of their Mira sample bear periods in excess of 400 days, a regime beyond our sample. Further, a likely candidate for such a process is that of the C-rich Mira discussed in \citet{Feast_2013} with a period of 551 days. Thus, we believe such effects are not at play in the age estimation of our longest period Miras.}

\section{Conclusions}

We take advantage of the, to date, underutilised sample of Miras provided by $Gaia$ DR2. Selecting O-rich Miras based on their distinct infrared colour and considerable photometric amplitude of variability, we map the Mira population through the Galactic disk and into the bulge. We exploit the correlation between Mira pulsation period with stellar age/chemistry to date our Mira sample and slice the Galaxy chronologically. Spanning ages of $\sim 5-10$ Gyr, we can observe how the structure of the disk and bulge/bar has evolved over the Milky Way's lifetime.

 We find the old, metal poor disk to be stubby; its vertical extent is large and is constricted radially. On increasing period, the disk settles into a thinner and radially more extensive formation. Modelling the disk as a simple double exponential, we see a progression in scale length from $\sim 3.8$ kpc to $\sim 4.5$ kpc for periods ranging from $100$ to $400$ days. Similarly, the vertical scale height of the disk plummets from $\sim 1.0$ kpc at low periods to $\sim 0.5$ kpc for our longer period Miras. This is entirely complementary to the findings of \citet{Bovy_2012} who studied the stellar disk profile as a function of chemical abundance using a large sample of G-type dwarfs from the SDSS-SEGUE survey. They find the scale height of the disk to ascend from $\sim 200$ pc to $\sim 1$ kpc on increasing stellar age, as indicated by their metallicity and $\left [ \alpha/\textup{Fe} \right ]$ abundance. The radial scale length of the disk decreases from $ > 4.5$ kpc down to $\sim 2$ kpc also on increasing stellar age. The structure we see in the disk could be understood in terms of the early Galaxy being turbulent and dynamically active, heating the early star forming disk. As the Galaxy became more quiescent, the younger stellar populations were born out of progressively cooler and more rotationally supported gas, thinning and extending the stellar disk \citep[see e.g.][]{Bournaud_2009,Bird_2013}.
 
 Within the disk, we see evidence of disruption and heating. A large ring-like structure is observed in the youngest of disk Miras, coincident with the expected location of the Outer Lindblad Resonance and indicative of secular evolutionary processes, mixing and restructuring the disk. Miras are observed to lie scattered from the plane, extending up to beyond $3$ kpc likely due to vertical perturbations induced in the disk over their lifetime.
 
 In the central regions of the Galaxy, we find the younger/metal-richer Miras to clearly inhabit an inclined bar-like structure. Fitting a boxy-bulge model to these Miras yields a viewing angle of $\sim 21^\circ$. The ancient Miras show little evidence for inclination, suggestive that they are a spatially distinct component from the bulge/bar. This is in accord with the findings of \citet{Dekany_2013} who see the bulge RR Lyrae to lie detached from the bar with little evidence for an inclined population. Satisfyingly, the age of our eldest Miras are comparable to those of RR Lyrae at $\sim 9-10$ Gyr. This distinction is also seen in the simulations of \citet{Debattista_2017} in which the disk is seeded by co-spatial populations of stars with differing velocity dispersion profiles. Bar formation occurs after $\sim 2$ Gyr, subsequently buckling into a bulge where the older, hotter stars form a boxy shape and the younger, cooler stars more easily trapped into X-shape orbits. When they studied the morphology of their oldest bulge stars, $> 9.5$ Gyr, they saw them to be decoupled from the bar structure. Thus, it is reasonable to interpret the RR Lyrae distribution of \citet{Dekany_2013} and our ancient Miras to represent those stars that were born prior to bar formation, and were kinematically hot enough to avoid entrapment by the bar potential. Given that the lowest period bin to demonstrate a bulge/bar structure occurs at $250-300$ days, we postulate that the bar must have buckled by $\sim 8-9$ Gyr ago, given the period age relations of \citet{Wyatt_1983} and \citet{Feast_2009}. 
 
 We find a bulge/bar half-length of $\sim 1.7-2$ kpc from our Miras, with an indication that the bulge/bar width and height decreases on stellar age. For the two period bins that clearly show the Miras to be bar-like, we see tentative evidence for an increase in the bar length. This is consistent with the trends seen by \citet{Wegg_2015}, who see the bar extending beyond the bulge to become longer and thinner for younger Red Clump stellar populations. Again our results are consistent with the findings of \citet{Debattista_2017} who see the younger stars to constitute a narrower and thinner bulge/bar.
 
 When comparing the spatial distributions of the oldest and youngest Miras in the inner Galaxy we see a clear distinction between the two. The old/metal-poor Miras dominate centrally, whilst the young/metal-rich Miras exhibit a boxy/peanut like morphology with a characteristic X-shape. This is consistent with the notion of 'kinematic fractionation' posited by \citet{Debattista_2017} wherein the younger, initially cooler stars couple more strongly to the buckling bar, exhibiting a more pronounced peanut-shape than the older stars.

By making use of the $Gaia$ DR2 LPV data set, we have shown it is possible to cleanly select a large sample of O-rich Miras and assign them accurate distances. Owing to their impressive brightness in the near infrared regime, we are able to trace the Miras right across the disk and through the bulge. Seizing on the correlation between their pulsation period and stellar age/chemistry, we can, for the first time, chronologically slice through the Galactic disk and bulge at once. The Miras are not the only variable star to be utilised in mapping out the Milky Way. Cepheid variables and RR Lyrae have long been used to do so but the Miras are not limited to a single age bracket nor do they primarily trace a single structure as the Cepheids do in the spiral arms. Many studies have made excellent use of the vast Red Clump Giant populations but these too lack the ability to act as chronometers in tracing the bulge/bar structure. Looking further afield, many Miras have already been resolved in M31 \citep[see e.g.][]{An_2004}, so it seems their ability to chronologically dissect a galaxy is not limited to the Milky Way. 

\section*{Acknowledgements}
JG thanks the Science and Technology Facilities Council (STFC) of the United Kingdom for a research grant.




\bibliographystyle{mnras}
\bibliography{bibliography} 



\appendix

\section{Residuals of softened exponential disk model}
\begin{figure*}
\centering
	\includegraphics[width=\textwidth]{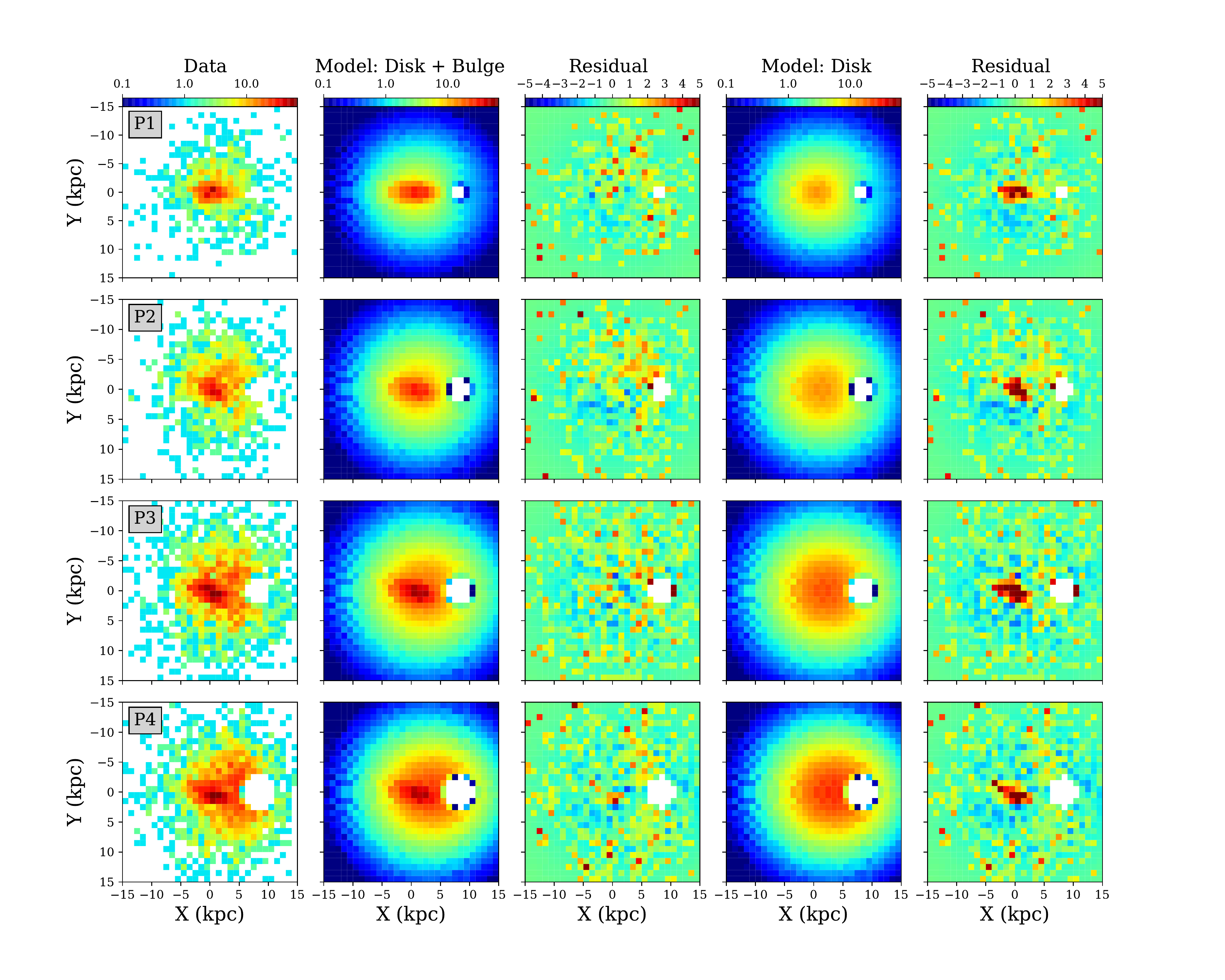}
    \caption[width=\textwidth]{This figure is equivalent to Fig.~\ref{fig:XY_residuals} but for our softened exponential disk model (M2). Negligible difference is observed in the Disk + Bar residual between that shown here and those of M1. The residuals of the disk only fit show a more pronounced central bulge over-density, owing to the flattening of the disk model in this region.  }
\label{fig:XY_res_m2a}
\end{figure*}
\begin{figure*}
\centering
	\includegraphics[width=\textwidth]{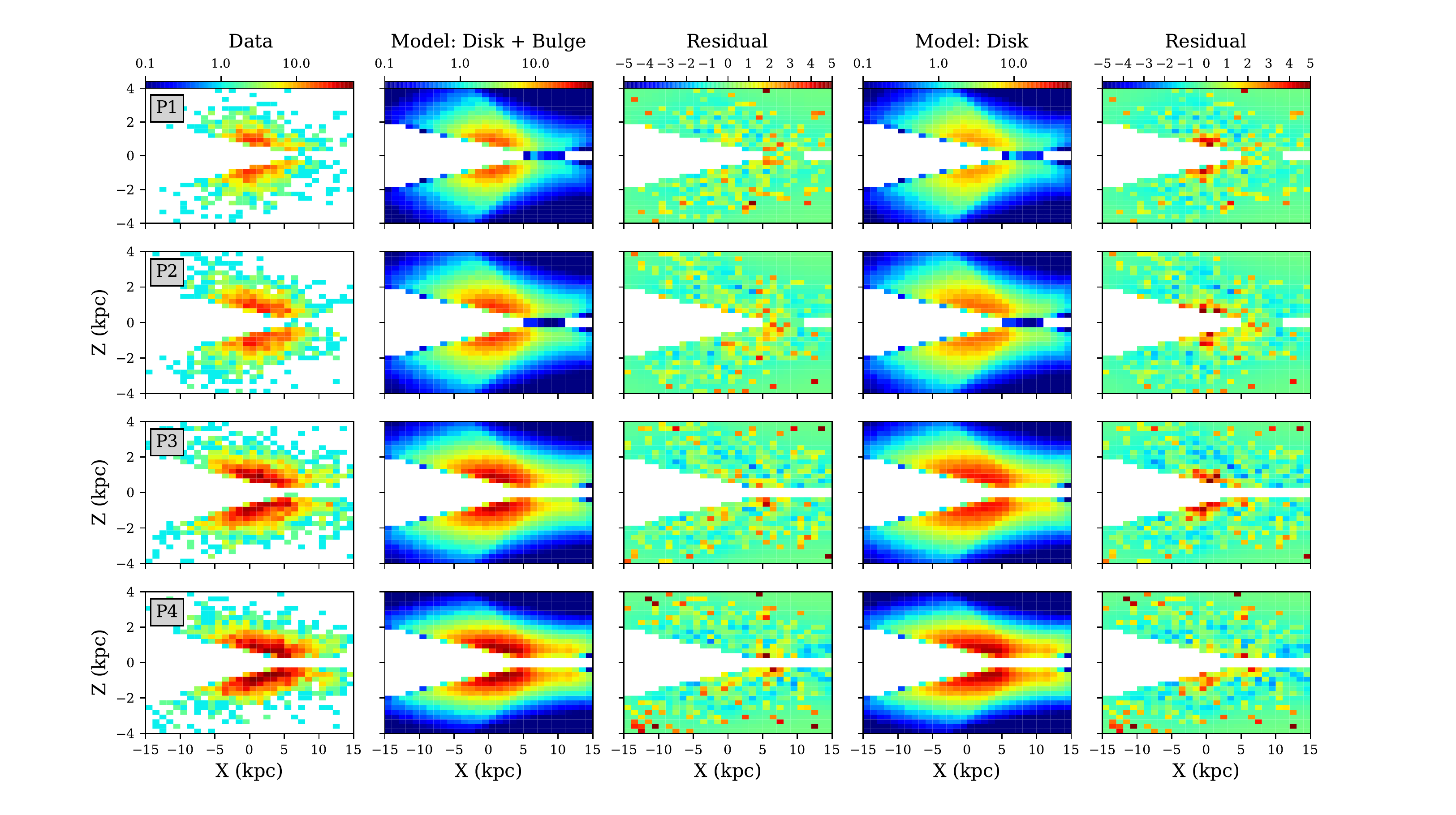}
    \caption[width=\textwidth]{Galactocentric X-Z projection of M2 showing good residual agreement with M1 in the case of the Disk + Bulge fit.}
\label{fig:XY_res_m2b}
\end{figure*}
\begin{figure*}
\centering
	\includegraphics[width=\textwidth]{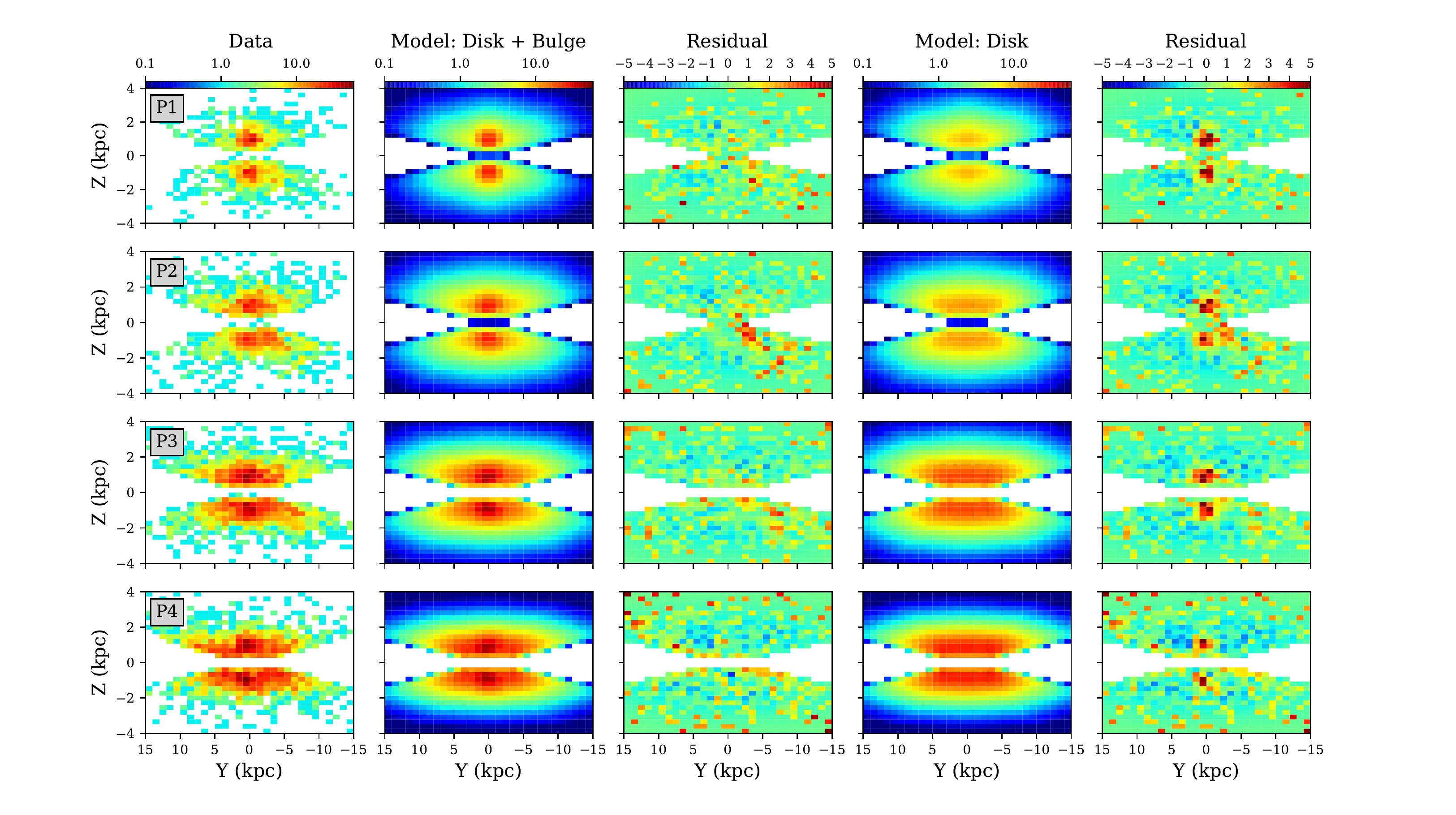}
    \caption[width=\textwidth]{Galactocentric Y-Z projection of our M2 model and residuals. Again we see little difference in the residual distribution when compared to M1.}
\label{fig:XY_res_m2c}
\end{figure*}
%


\bsp	
\label{lastpage}
\end{document}